\documentclass{article} % For LaTeX2e
\usepackage{iclr2027_conference,times}

\usepackage{amsmath,amsfonts,bm}

\def\eqref#1{equation~\ref{#1}}
\def\1{\bm{1}}

\DeclareMathAlphabet{\mathsfit}{\encodingdefault}{\sfdefault}{m}{sl}
\SetMathAlphabet{\mathsfit}{bold}{\encodingdefault}{\sfdefault}{bx}{n}

\usepackage{hyperref}
\usepackage{url}
\usepackage{booktabs}

\usepackage{array,tabularx}
\newcommand{\PGTableStyle}{%
  \small
  \renewcommand{\arraystretch}{1.14}%
  \setlength{\tabcolsep}{5pt}%
  \ifdefined\hypersetup\hypersetup{pdfborder={0 0 0}}\fi
}

\newcommand{\PGPair}[2]{%
  \makebox[2.5em][r]{#1}\,/\,\makebox[2.5em][r]{#2}%
}

\usepackage{tikz}
\usetikzlibrary{arrows.meta,positioning,calc}
\usepackage{amssymb}

\title{ProofGap: Benchmarking Step-Level Formal Reasoning with Local Obligations Derived from Natural-Language Solutions}

\author{%
\begin{minipage}[t]{\dimexpr\textwidth-2\tabcolsep\relax}
\centering
Lihan Xie\textsuperscript{1,2}\quad
Zhicheng Hui\textsuperscript{1}\quad
Yingjun Lan\textsuperscript{1}\\[3pt]
Zhehao Li\textsuperscript{1}\quad
Xingzhi Qi\textsuperscript{1}\quad
Siyue Huang\textsuperscript{1}\quad
Jirui Liu\textsuperscript{1}\\[3pt]
Chuxiao Zeng\textsuperscript{1}\quad
Bohan Zhao\textsuperscript{1}\quad
Qinxiang Cao\textsuperscript{1,}\thanks{Corresponding author.}\\[6pt]
\normalfont\small
\textsuperscript{1}Shanghai Jiao Tong University\\[2pt]
\textsuperscript{2}Shanghai Innovation Institute
\end{minipage}%
}

\iclrfinalcopy % Show author names and disable the review line ruler.
\hypersetup{hidelinks} % Keep preprint links clickable without colored boxes.

\usepackage{mathtools,listings,needspace,placeins}
\definecolor{PGAppendixCodeBg}{RGB}{248,248,248}
\definecolor{PGAppendixCodeRule}{RGB}{205,205,205}
\lstdefinestyle{PGAppendix}{
  basicstyle=\small\ttfamily,
  backgroundcolor=\color{PGAppendixCodeBg},
  frame=single,rulecolor=\color{PGAppendixCodeRule},framerule=0.35pt,
  xleftmargin=6pt,xrightmargin=6pt,framesep=5pt,
  aboveskip=8pt,belowskip=8pt,
  breaklines=true,breakatwhitespace=true,breakindent=1.5em,
  columns=fullflexible,keepspaces=true,showstringspaces=false,tabsize=2
}
\newcommand{\PGAppendixTableStyle}{%
  \small\renewcommand{\arraystretch}{1.13}\setlength{\tabcolsep}{4pt}%
}
\newcommand{\PGAppendixPair}[2]{%
  \makebox[2.5em][r]{#1}\,/\,\makebox[2.5em][r]{#2}%
}
\newcommand{\rnfl}{Relaxed NFL}
\newcommand{\cnfl}{Core NFL}
\newcommand{\term}{\mathsf{Term}}
\newcommand{\method}{\mathsf{Method}}
\newcommand{\pstate}{\mathsf{PState}}
\newcommand{\gap}{\mathsf{Gap}}
\newcommand{\pgg}{\mathsf{PGG}}
\newcommand{\checkgap}{\mathsf{CheckGap}}
\newcommand{\checkprogram}{\mathsf{CheckProgram}}
\newcommand{\code}[1]{\texttt{#1}}

\begin{document}

\maketitle
\lhead{Preprint}

\begin{abstract}
% The abstract paragraph should be indented 1/2~inch (3~picas) on both left and
% right-hand margins. Use 10~point type, with a vertical spacing of 11~points.
% The word \textsc{Abstract} must be centered, in small caps, and in point size 12. Two
% line spaces precede the abstract. The abstract must be limited to one
% paragraph.
Existing formal mathematics benchmarks, such as miniF2F, ProofNet, and PutnamBench, primarily evaluate models on constructing complete formal proofs for challenging problems. Because success is measured at the theorem level, these benchmarks offer limited insight into models’ step-level formal reasoning. Evaluating this capability separately enables finer-grained diagnosis of model limitations than theorem-level evaluation alone.

To fill this evaluation gap, we introduce ProofGap, a fine-grained benchmark for step-level formal reasoning. ProofGap is constructed through a natural-language proof-processing pipeline that decomposes each reasoning step into one or more aligned proof gaps. Applying this pipeline to natural-language solutions to 3,015 exercises in B. P. Demidovich’s Problems in Mathematical Analysis yields 26,116 gaps. The benchmark focuses on mathematical analysis, a domain that remains challenging for current models. By supplying the local context and target explicitly, gap completion isolates local formal proof construction from end-to-end proof composition, enabling more precise localization of model failures. Natural-language solutions serve as the provenance of these obligations, while the benchmark task itself starts from an already formalized local context and goal. Beyond benchmarking, the same pipeline may support future proof-verification systems, provided that semantic translation and sequential proof composition are handled reliably.
\end{abstract}

\section{Introduction}
Mathematical reasoning provides a demanding test of whether large language models can perform rigorous, multi-step inference. Although recent models have achieved impressive performance on natural-language mathematics, their proofs may still contain subtle logical errors, omitted assumptions, or unsupported intermediate claims that are difficult to detect reliably. Formal systems such as Lean, Rocq, and Isabelle offer a complementary approach by expressing mathematical arguments in precise languages and mechanically checking the resulting proofs \cite{moura2021lean, bertot2013interactive, nipkow2002isabelle}. Building on this verifiable interface, powerful formal theorem-proving systems such as DeepSeek-Prover \cite{ren2025deepseekproverv2advancingformalmathematical} and AlphaProof \cite{hubert2026olympiad} have emerged. In parallel, benchmarks such as miniF2F, ProofNet, and PutnamBench have provided standardized environments for evaluating statement formalization and formal proof construction across challenging mathematical problems. Despite differences in their specific tasks \cite{zheng2021minif2f, tsoukalas2024putnambench, azerbayev2023proofnet}, these benchmarks largely take the complete theorem as the basic unit of evaluation. 

Yet in many practical settings, what is needed is not formal reasoning over a complete proof, but fine-grained feedback on individual reasoning steps. During natural-language proof generation, a critical step can be subjected to local formal verification as soon as it is proposed, allowing unsupported inferences to be detected and revised before their errors propagate through the subsequent argument. Such feedback can also be used to filter candidate reasoning steps or rank alternative reasoning trajectories, allowing models to prioritize directions supported by formal evidence. SAFE exemplifies this use case: it formalizes individual model-generated reasoning steps as Lean statements and combines the resulting proof states with a process reward model to improve Best-of-N trajectory selection \cite{liu2025safe}.

Step-level formal feedback is also valuable during training. Prior work has shown that process supervision over intermediate reasoning steps can provide more effective training signals than outcome supervision based solely on final answers \cite{lightman2024let}. FoVer further uses formal verification tools to generate step-level error labels automatically, demonstrating that formal systems can provide machine-checkable training data for process reward models \cite{kamoi2026efficient}. Such feedback can supervise the reasoning process itself, rather than assigning credit solely according to whether the final answer is correct.

Both settings require formal feedback that is trustworthy, localized, and sufficiently inexpensive to obtain repeatedly. Whole-proof formalization and proof search provide only a coarse signal and can be computationally costly when applied to every candidate proof or sampled training trajectory. A failure at the whole-proof level is also difficult to interpret: the natural-language argument may contain a logical error, the autoformalization component may translate the proof goal incorrectly, or the formal proof-generation model may fail even when given an accurately formalized and provable goal. MA-ProofBench makes this ambiguity concrete in mathematical analysis. Even with expert-reviewed Lean statements, several models that produce many natural-language proofs judged to be fully correct nevertheless achieve near-zero formal proof success on the same problems, with Mathlib hallucinations and incomplete proofs among the dominant failure modes \cite{pu2026ma}. Thus, whole-proof verification provides only a coarse signal for applications that require localized and actionable formal feedback.

For formal feedback to be reliable and actionable, each verification task should correspond to an identifiable reasoning step in the source natural-language proof, rather than to an arbitrary fragment of a formal derivation. To this end, we introduce ProofGap, a fine-grained benchmark for step-level formal reasoning. Using a semantics-aware pipeline for processing natural-language proofs, we construct ProofGap from 3,015 exercises originating from
Demidovich's \textit{Problems in Mathematical Analysis} \cite{demidovich1969problems} and their Chinese reference solutions
compiled by Fei and Zhou \cite{fei2012demidovich}. 

ProofGap focuses on mathematical analysis not only because formal proof construction in this domain remains challenging for current models \cite{yu2025formalmath, pu2026ma}, but also because many local inferences in analysis solutions depend on domain restrictions, nonzero conditions, sign conditions, and convergence assumptions that are easily left implicit in natural-language arguments. These characteristics make mathematical analysis a particularly informative testbed for evaluating step-level formal reasoning. Moreover, the breadth and diversity of Demidovich’s exercise collection provide a suitable foundation for the systematic, large-scale construction of such localized formal reasoning tasks.

To construct ProofGap, we build on a verifiable auto-formalization pipeline based on the Relaxed and Core Natural Formal Languages. The pipeline first translates a natural-language solution into an intermediate representation that preserves its stepwise reasoning structure, and then resolves its ambiguous notation, problem-solving constructs, and implicit variable scopes through a verifiable elaboration procedure. From the resulting semantically defined proof representation, a proof-gap generator derives localized verification conditions that expose the assumptions available at each point and the goal to be established. ProofGap takes the discharge of these conditions as its benchmark task.

% todo：experiment describe
% As an initial baseline, \texttt{autosolve} closes 2,823 gaps, while a verifier-guided Codex workflow using \texttt{gpt-5.6-sol} produces checker-validated DSL proofs for 4,677 more, solving 7,500 of 26,116 gaps (28.7\%) in total.

Our main contributions are as follows:
\begin{itemize}
    \item We formulate step-level formal reasoning as a distinct evaluation problem in which models discharge local formal proof obligations derived from individual reasoning steps in textbook natural-language solutions.

    \item We construct ProofGap through a natural-language proof-processing procedure, obtaining 26,116 formal proof gaps from multi-step solutions to 3,015 exercises in mathematical analysis, and release both its original and Lean versions.\footnote{Project repository: \url{https://github.com/xielihan/ProofGap-Benchmark}}

    \item For the original version, we provide a lightweight checker and a verifier-guided agent workflow to produce executable, checker-validated proofs; for the Lean version, we provide ground-truth proofs for exercises and gaps.

    \item We systematically evaluate a range of models on ProofGap and conduct fine-grained analyses to identify their major limitations in formal reasoning for mathematical analysis.
\end{itemize}

By evaluating models on localized, mechanically checkable proof obligations, ProofGap provides a fine-grained benchmark for studying formal reasoning beyond theorem-level success. These obligations may also serve as components of future natural-language proof-verification systems.

\section{Related Work}

Research most closely related to ProofGap lies at the intersection of
formal mathematics benchmarks, step-level mathematical verification,
and formal reasoning in mathematical analysis.
Table~\ref{tab:benchmark-comparison} compares representative formal
mathematics benchmarks in terms of mathematical scope, target source,
and evaluation granularity.

\begin{table}[!hb]
\centering
\caption{
Comparison of formal mathematics benchmarks by mathematical scope,
target provenance, and evaluation granularity.
Granularity is defined relative to the source problem or reasoning
process; every evaluation target is nevertheless represented as a
standalone formal theorem. NL denotes natural language.
}
\label{tab:benchmark-comparison}
\vspace{4pt}
\PGTableStyle
\setlength{\tabcolsep}{4pt}
\renewcommand{\tabularxcolumn}[1]{m{#1}}
\begin{tabularx}{\linewidth}{@{}>{\raggedright\arraybackslash}m{.25\linewidth}>{\raggedright\arraybackslash}m{.17\linewidth}>{\raggedright\arraybackslash}X>{\raggedright\arraybackslash}m{.16\linewidth}r@{}}
\toprule
\textbf{Benchmark} &
  \textbf{Mathematical scope} &
  \textbf{Target source} &
  \textbf{Granularity} &
  \textbf{\# Targets} \\
  \midrule
miniF2F\newline\cite{zheng2021minif2f} &
Multi-domain &
Original problem &
Theorem-level &
488 \\
\addlinespace[2pt]
ProofNet\newline\cite{azerbayev2023proofnet} &
Multi-domain &
Original problem &
Theorem-level &
371 \\
\addlinespace[2pt]
PutnamBench\newline\cite{tsoukalas2024putnambench} &
Multi-domain &
Original problem &
Theorem-level &
640 \\
\addlinespace[2pt]
FormalMATH\newline\cite{yu2025formalmath} &
Multi-domain &
Original problem &
Theorem-level &
5,560 \\
\addlinespace[2pt]
MA-ProofBench\newline\cite{pu2026ma} &
Mathematical analysis &
Original problem &
Theorem-level &
200 \\
\addlinespace[2pt]
FormalML\newline\cite{yang2025formalml} &
Machine learning theory &
Formal proof subgoal &
Step-level &
4,937 \\
\addlinespace[2pt]
FormalStep\newline\cite{liu2025safe} &
Multi-domain &
Model-generated NL reasoning step &
Step-level &
30,809 \\
\addlinespace[2pt]
\textbf{ProofGap} &
Mathematical analysis &
Original textbook NL solution step &
Step-level &
\textbf{26,116} \\
\bottomrule
\end{tabularx}
\end{table}

\paragraph{Formal Mathematics Benchmarks.}
Existing benchmarks primarily construct formal evaluation targets from
original mathematical problems. Representative benchmarks include
miniF2F \cite{zheng2021minif2f}, ProofNet
\cite{azerbayev2023proofnet}, PutnamBench
\cite{tsoukalas2024putnambench}, and FormalMATH
\cite{yu2025formalmath}.
Despite differences in problem source, domain, and difficulty, these
benchmarks generally derive their formal targets directly from the
original problem statements and report whether models successfully
prove them at the problem level.
This setting is well suited to evaluating end-to-end formal proof
construction, but provides limited insight into whether a model can
formally justify a particular inference in the corresponding
natural-language argument.
Notably, although ProofNet provides natural-language proofs, it does not
decompose their individual reasoning steps into aligned formal proof
obligations.

\paragraph{Step-Level Mathematical Verification.}
Step-level reasoning has been studied through process-supervision
datasets that label intermediate natural-language steps
\citep{lightman2024let,wang2024math}, and through formally grounded
settings such as FoVer, FormalML, and SAFE/FormalStep
\citep{kamoi2026efficient,yang2025formalml,liu2025safe}.
ProofGap differs from these works in the source of its formal targets
and in its alignment criterion.
FormalML extracts subgoals from existing Lean developments, while
FormalStep autoformalizes model-generated reasoning trajectories, whose
steps may be incorrect, redundant, or self-correcting. In contrast,
ProofGap decomposes coherent textbook reference solutions into local
proof obligations aligned with the logical structure of complete
natural-language arguments.

\paragraph{Formal Reasoning in Mathematical Analysis.}
Mathematical analysis is particularly challenging for formal reasoning:
local inferences often depend on domain restrictions, nonzeroness,
differentiability, integrability, convergence assumptions, and
library-specific representations of analytic concepts.
Recent evaluations confirm this difficulty: FormalMATH reports
substantially lower formal proof success on calculus than on algebra,
and MA-ProofBench finds that models still struggle with mathematical
analysis even when they perform much better on the corresponding
natural-language problems. Precisely because end-to-end success remains low in this domain, fine-grained diagnosis is important: theorem-level failure alone cannot
distinguish whether a model fails on analytic side conditions, local
algebraic transformations, limit arguments, or the global proof plan.
However, existing analysis benchmarks remain problem-level, constructing
one formal target for each original problem.
ProofGap complements them by processing reference solutions to 3,015
Demidovich exercises into 26,116 step-level formal targets, enabling
localized evaluation of formal reasoning in mathematical analysis.

\section{The ProofGap Benchmark}

\subsection{Task Definition}

A ProofGap instance is represented as a triple
\begin{equation}
(\Gamma,g,m),
\end{equation}
where $\Gamma$ is the well-scoped local context available at the
current proof position, $g$ is the target proposition, and $m$ is an
optional strategy annotation preserved from the source solution. These
components correspond to the \textit{ASSUM}, \textit{GOAL}, and
\textit{METHOD} fields in the released data.

Given $\Gamma$ and $g$, a model must generate a formal derivation
accepted by the designated proof checker, thereby establishing
\begin{equation}
\Gamma\vdash g.
\end{equation}
The annotation $m$ is optional metadata rather than part of the proof
obligation: it neither affects checker acceptance nor is provided to
models in the reported evaluations. We release it to support future
analyses of proof strategies and strategy-conditioned proving.

Each gap is evaluated independently. Intermediate propositions
established earlier in the source solution are included in $\Gamma$ and
treated as available premises. The source natural-language step is used during benchmark construction and traceability analysis, but is not part of the model input. ProofGap therefore evaluates local formal proof completion over obligations derived from natural-language solutions. It does not directly evaluate natural-language autoformalization or the detection of invalid natural-language steps. Moreover, because intermediate propositions from earlier steps are supplied as premises, the task does not evaluate end-to-end proof composition.

\begin{figure*}[t]
\centering
\includegraphics[width=\textwidth]{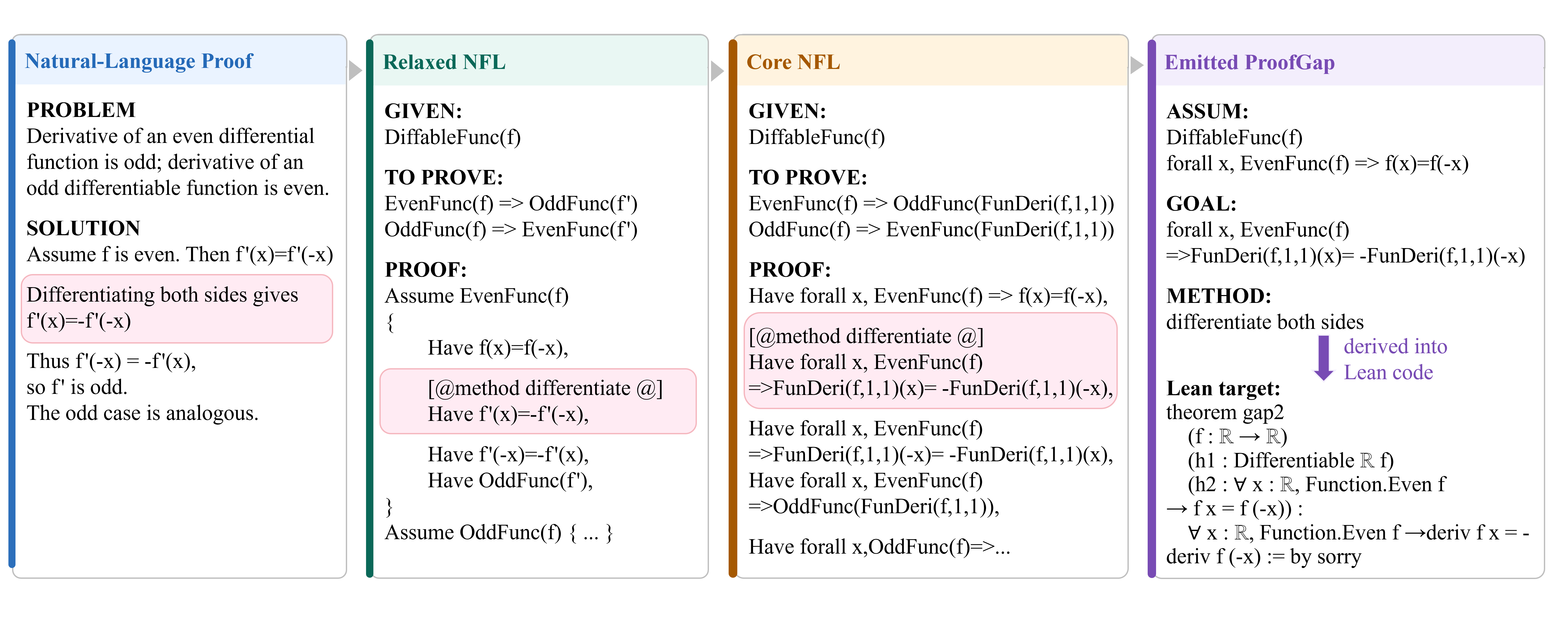}
\caption{Construction pipeline from a boxed natural-language reasoning step to a local ProofGap instance. Relaxed NFL preserves the original proof structure and method annotation, while Core NFL makes the implicit universal variables and assumption scopes explicit and normalizes \(f'\) as \textit{FunDeri(f,1,1)}, where \textit{FunDeri(f,1,1)} denotes the first-order derivative of function \(f\) with respect to its first formal argument. The last column shows the emitted ProofGap instance and its corresponding Lean target.}
\label{fig:proofgap-construction}
\end{figure*}

\subsection{Benchmark Construction}

We construct ProofGap from exercises in B. P. Demidovich's
\emph{Problems in Mathematical Analysis} and their Chinese reference
solutions compiled by Fei and Zhou. The corpus covers major topics in
mathematical analysis. Before formalization, we remove problems that
depend on figures or unavailable external information, split independent
subquestions, and normalize problem--solution boundaries.

Figure~\ref{fig:proofgap-construction} illustrates the construction
pipeline. In the first stage, a prompt-based translator converts each
problem and solution into Relaxed Natural Formal Language (Relaxed NFL),
a parser-readable intermediate representation that preserves the
conditions, goal, proof structure, and method annotations of the source.
The generated representation then undergoes parser validation and
representation-consistency screening, which checks the preservation of
the main assumptions, goal, and proof steps and rejects unsupported
intermediate conclusions. Detailed screening criteria are provided in Appendix~\ref{app:relaxed-nfl}.

In the second stage, the Relaxed NFL AST is elaborated into Core NFL, a
normalized representation with explicit proof contexts and variable
bindings. The elaboration decomposes compound propositions,
disambiguates context-dependent notation, canonicalizes mathematical
operations, and makes remaining free variables explicit. Most
transformations are deterministic; unresolved notation or scope
ambiguities are handled using constrained LLM heuristics followed by
programmatic validation. The resulting Core NFL representation must
pass syntax and binding-completeness checks.

In the third stage, the proof-gap generator traverses the Core NFL AST
while maintaining the current proof state $\Gamma\vdash g$. When a step
introduces an intermediate proposition $P$, the generator emits the
obligation $\Gamma\vdash P$ and adds $P$ to the context used to construct
subsequent proof states, without requiring the preceding gap to be
solved. Nested and backward proof steps are handled by generating the
corresponding local obligations. Each gap retains a trace link to its
source step and Core NFL node; a single source step may produce multiple
gaps when compound or nested reasoning is decomposed. Formal generation rules are provided in Appendix~\ref{app:gap-generation}.

The generator first emits each local obligation in the benchmark-native
ProofGap representation. To support evaluation with existing formal
proof assistants, every generated gap is then translated into a Lean
proof obligation using GPT-5.6-sol \cite{openai2026gpt56}. Local declarations and assumptions are mapped to Lean binders and theorem premises, while the gap goal is mapped to the
theorem conclusion. Each Lean theorem statement is paired with a proof verified by Lean prover. The translations are audited through a human-in-the-loop cross-review
protocol. Multiple independently instantiated LLM reviewers compare
each formalized obligation with the corresponding source problem and
solution step. Any case flagged by at least one reviewer for a
potential translation error or semantic ambiguity is escalated to
human adjudication and retained only after its semantic fidelity has
been confirmed.

\subsection{Statistics and Coverage}

The final ProofGap benchmark contains 3,015 mathematical-analysis
exercises and 26,116 proof gaps. Each exercise yields 8.66 gaps on
average, with a median of 7. Each gap retains 8.33 assumptions on
average, with a median of 6. These statistics show that a complete
natural-language solution is typically decomposed into multiple local
proof obligations, while each obligation preserves several facts
available at the corresponding step of the solution.

We group the exercises into seven topics according to the organization
of the source material. As shown in Table~\ref{tab:proofgap-topics},
ProofGap covers single-variable functions, limits, and continuity;
single-variable differentiation; indefinite and definite integrals;
series; multivariable differentiation; parameter integrals; and
multiple, line, and surface integrals. Parameter integrals are treated
as a separate category because their proof obligations frequently
involve parameter-dependent limits, convergence conditions, and
interchanges between limiting operations and integration.

\suppressfloats[t]
\begin{table}[!hb]
\centering
\caption{Topic coverage of ProofGap.}
\label{tab:proofgap-topics}
\vspace{4pt}
\PGTableStyle
\begin{tabular*}{\linewidth}{@{\extracolsep{\fill}}lrr@{}}
\toprule
\textbf{Topic} & \textbf{Exercises} & \textbf{Gaps} \\
\midrule
Functions \& Limits & 529 & 4,033 \\
Single-Var. Differentiation & 640 & 5,019 \\
Indefinite and Definite Integrals & 620 & 4,795 \\
Infinite Series & 455 & 4,986 \\
Multivariable Differentiation & 260 & 2,380 \\
Parameter Integrals    & 100 & 1,082 \\
Multiple, Line, and Surface Integrals & 411 & 3,821 \\
\bottomrule
\end{tabular*}
\end{table}

The topic distribution is not uniform, reflecting the composition of
the source material. Nevertheless, every topic contains at least 100
exercises and more than 1,000 proof gaps. The benchmark therefore
supports both aggregate evaluation and topic-level analysis across
major components of undergraduate mathematical analysis.

\section{A Lightweight Proof-Checking Framework for ProofGap}
\label{sec:lightweight-framework}

For the original version of ProofGap, we provide a lightweight proof-checking
interface. Given a local context $\Gamma$ and goal $g$, a model emits a short
executable program in an 18-command domain-specific language (DSL), and a
deterministic checker replays it against the evolving proof state. A program
passes the checker only when every generated goal closes, yielding a
replayable certificate.

The framework provides deterministic solving through \textit{autosolve} and
access to a 514-entry theorem library. Unsupported solver inputs return
\emph{unknown}, and theorem applications must discharge all premises and side
conditions. A verifier-guided agent uses checker feedback to revise candidate
programs; accepted certificates additionally undergo admit-free root replay
and semantic audit. Appendix~\ref{app:framework-details} gives the command,
checker, and solver specifications, while Appendix~\ref{app:agent-workflow}
and Figure~\ref{fig:agent-baseline} describe the agent workflow.

\paragraph{Baseline results.}
Treating \textit{autosolve} as a lightweight automated theorem prover, we first
apply this DSL command to all 26,116 ProofGap instances and directly close
2,838 gaps (10.87\%).  We then run our verifier-guided agent workflow on the remaining
instances, using Codex as the agent and GPT-5.6-sol as its underlying
model; the workflow synthesizes checker-validated DSL proofs for an additional
4,889 gaps.  Together, the two stages solve 7,727 gaps (29.59\%).  Beyond model
scoring, each accepted DSL program is an executable ground-truth solution for
the corresponding gap in the original version of the benchmark.

\section{Experiment}
\subsection{Research Questions}
Our experiments on the Lean version of ProofGap are organized
around four research questions.
\begin{itemize}
  \item \textbf{RQ1:}
  To what extent can the evaluated models solve the step-level
  formal proof obligations in ProofGap?

  \item \textbf{RQ2:}
  Under a fixed per-target sampling budget, how does performance on step-level gaps compare with performance on their corresponding complete exercise theorems?

    \item \textbf{RQ3:}
    How do specialized formal provers and general-purpose language
    models compare on the balanced ProofGap-280 subset?

  \item \textbf{RQ4:}
  What additional diagnostic information does step-level evaluation
  provide beyond the success or failure of complete theorem proving?
\end{itemize}

\subsection{Experimental Setup}

\noindent\textbf{Evaluation sets.}
The complete Lean evaluation set contains 26,116 step-level gaps from
3,015 exercise instances, together with their corresponding 3,015
exercise theorems. Every target is accompanied by a Lean-verified
reference proof that is withheld from the evaluated model. For
cross-model comparison, we construct ProofGap-280 by sampling 10 gaps
from each of the 28 strata defined by seven mathematical topics and
four length categories. The selected gaps come from 280 distinct
parent exercises, yielding 280 paired gap--exercise targets. Further details on the construction of ProofGap-280, together with
model performance stratified by mathematical topic and proof length,
are provided in Appendices~\ref{app:proofgap-280} and~\ref{app:proofgap-280-breakdown}.

\noindent\textbf{Models.}
We evaluate Goedel-Prover-V2-8B \cite{lin2026goedelproverv}, DeepSeek-Prover-V2-7B, and
Kimina-Prover-Distill-8B \cite{wang2025kiminaproverpreviewlargeformal} on the full paired evaluation set and
ProofGap-280. On ProofGap-280, we additionally evaluate
Claude-Opus-4.8 \cite{anthropic2026claudeopus48}, GPT-5.6-sol, Kimi-K3 \cite{kimiteam2026kimik3openfrontier}, GLM-5.2 \cite{glm5team2026glm5vibecodingagentic}, and
DeepSeek-V4-Pro \cite{deepseekai2026deepseekv4highlyefficientmilliontoken}. Model configurations are provided in Appendix~\ref{app:model-budget}.

\noindent\textbf{Verification and metrics.}
We use a budget of eight candidate proofs per target, each capped at 8,192
tokens, and evaluate proof correctness with Lean. We report Pass@$k$ using
the standard estimator of \citet{chen2021evaluating}. Gap-level scores assign
equal weight to every gap (micro averaging). For exercise-level diagnosis at
Pass@8, we first compute the fraction of solved gaps within each exercise,
then average equally across exercises (macro coverage). Formal definitions
are given in
Appendix~\ref{app:evaluation-metrics}; full-set Pass@1/3/8 results are provided
in Appendix~\ref{app:full-passk}.

\subsection{Results}

\suppressfloats[t]
\begin{table}[!htbp]
\centering
\caption{
Pass@8 (\%) on all 26,116 proof gaps, the 23,191 gaps not solved by
the evaluated tactic portfolio, and the 3,015 corresponding exercise
theorems.
}
\label{tab:full-benchmark-results}
\vspace{4pt}
\PGTableStyle
\begin{tabular*}{\linewidth}{@{\extracolsep{\fill}}lrrr@{}}
\toprule
\textbf{Model}
& \shortstack{\textbf{All}\\\textbf{gaps}}
& \shortstack{\textbf{Tactic-unsolved}\\\textbf{gaps}}
& \shortstack{\textbf{Exercise}\\\textbf{theorems}} \\
\midrule
Goedel-V2-8B
& 35.39 & 27.48 & 4.28 \\
DeepSeek-Prover-7B
& 33.27 & 25.16 & 3.32 \\
Kimina-Distill-8B
& 31.99 & 23.57 & 2.92 \\
\bottomrule
\end{tabular*}
\end{table}

\begin{table}[t]
\centering
\caption{
Topic-level Pass@8 (\%). Each entry reports
\emph{gap/exercise-theorem} performance.
}
\label{tab:topic-results}
\vspace{4pt}
\PGTableStyle
\begin{tabular*}{\linewidth}{@{\extracolsep{\fill}}lccc@{}}
\toprule
\textbf{Topic}
& \textbf{Goedel}
& \textbf{DeepSeek}
& \textbf{Kimina} \\
\midrule
Functions, Limits, and Continuity
& \PGPair{44.41}{8.88}
& \PGPair{40.32}{7.56}
& \PGPair{39.10}{7.37} \\
\addlinespace[2pt]
Single-Variable Differentiation
& \PGPair{35.11}{4.69}
& \PGPair{32.72}{2.50}
& \PGPair{32.12}{2.97} \\
\addlinespace[2pt]
Indefinite and Definite Integrals
& \PGPair{32.43}{1.77}
& \PGPair{30.57}{1.45}
& \PGPair{28.59}{0.16} \\
\addlinespace[2pt]
Series
& \PGPair{31.09}{3.52}
& \PGPair{29.58}{3.08}
& \PGPair{27.20}{2.86} \\
\addlinespace[2pt]
Multivariable Differentiation
& \PGPair{40.21}{3.08}
& \PGPair{38.24}{2.69}
& \PGPair{37.73}{1.92} \\
\addlinespace[2pt]
Parameter Integrals
& \PGPair{20.43}{1.00}
& \PGPair{19.50}{0.00}
& \PGPair{18.21}{0.00} \\
\addlinespace[2pt]
Multiple, Line, and Surface Integrals
& \PGPair{36.82}{3.89}
& \PGPair{35.57}{3.41}
& \PGPair{35.15}{2.68} \\
\bottomrule
\end{tabular*}
\end{table}

\begin{table}[t]
\centering
\caption{
Mean exercise-level gap coverage, reported over all exercises and
conditioned on whether the corresponding complete theorem is solved
at Pass@8.
}
\label{tab:coverage-by-theorem-outcome}
\vspace{4pt}
\PGTableStyle
\begin{tabular*}{\linewidth}{@{\extracolsep{\fill}}lrrr@{}}
\toprule
& \multicolumn{3}{c}{\textbf{Mean gap coverage (\%)}} \\
\cmidrule(lr){2-4}
\textbf{Model}
& \shortstack{\textbf{All}\\\textbf{exercises}}
& \shortstack{\textbf{Theorem}\\\textbf{solved}}
& \shortstack{\textbf{Theorem}\\\textbf{failed}} \\
\midrule
Goedel
& 31.15 & 75.07 & 29.19 \\
DeepSeek
& 29.12 & 75.94 & 27.51 \\
Kimina
& 27.63 & 79.12 & 26.08 \\
\bottomrule
\end{tabular*}
\end{table}

\begin{table}[t]
\centering
\caption{
Local progress among exercises whose complete theorem is not solved
at Pass@8. All values are percentages of theorem-failed exercises.
}
\label{tab:failed-theorem-progress}
\vspace{4pt}
\PGTableStyle
\begin{tabular*}{\linewidth}{@{\extracolsep{\fill}}lrrr@{}}
\toprule
\textbf{Model}
& \shortstack{\textbf{Any gap}\\\textbf{solved}}
& \shortstack{\textbf{At least}\\\textbf{half}}
& \shortstack{\textbf{At least}\\\textbf{75\%}} \\
\midrule
Goedel
& 78.03 & 22.59 & 4.99 \\
DeepSeek
& 75.88 & 20.24 & 4.46 \\
Kimina
& 72.39 & 18.79 & 4.68 \\
\bottomrule
\end{tabular*}
\end{table}

\begin{table}[t]
\centering
\caption{
Pass@$k$ results (\%) on ProofGap-280 and its 280 corresponding
parent exercises. Each entry reports
\textit{proof-gap / parent-exercise} performance under an
8,192-token generation budget.
}
\label{tab:proofgap-280-results}
\vspace{4pt}
\PGTableStyle
\begin{tabular*}{\linewidth}{@{\extracolsep{\fill}}lccc@{}}
\toprule
& \multicolumn{3}{c}{\textbf{Proof gap / parent exercise}} \\
\cmidrule(lr){2-4}
\textbf{Model}
& \textbf{Pass@1}
& \textbf{Pass@3}
& \textbf{Pass@8} \\
\midrule

\multicolumn{4}{@{}l}{\textbf{\textit{Specialized Formal Provers}}} \\
\addlinespace[1pt]
Goedel-V2-8B
& \PGPair{29.74}{3.97}
& \PGPair{36.35}{5.64}
& \PGPair{39.29}{7.14} \\

DeepSeek-Prover-7B
& \PGPair{30.28}{4.06}
& \PGPair{36.77}{5.33}
& \PGPair{39.64}{5.71} \\

Kimina-Distill-8B
& \PGPair{30.15}{2.41}
& \PGPair{34.69}{3.42}
& \PGPair{35.71}{4.29} \\

\midrule
\multicolumn{4}{@{}l}{\textbf{\textit{Open-Weight General Models}}} \\
\addlinespace[1pt]
Kimi-K3
& \PGPair{70.71}{7.14}
& \PGPair{76.79}{8.21}
& \PGPair{81.07}{8.93} \\

DeepSeek-V4-Pro
& \PGPair{52.50}{2.86}
& \PGPair{62.14}{4.29}
& \PGPair{67.50}{5.00} \\

GLM-5.2
& \PGPair{47.86}{1.79}
& \PGPair{61.43}{2.50}
& \PGPair{65.00}{3.21} \\

\midrule
\multicolumn{4}{@{}l}{\textbf{\textit{Proprietary General Models}}} \\
\addlinespace[1pt]
Claude-Opus-4.8
& \PGPair{77.50}{8.21}
& \PGPair{83.93}{13.21}
& \PGPair{87.86}{13.93} \\

GPT-5.6-sol
& \PGPair{79.29}{12.99}
& \PGPair{84.97}{17.61}
& \PGPair{87.86}{21.43} \\
\bottomrule
\end{tabular*}
\end{table}

\subsubsection{Full-Set Solvability (RQ1)}

On the complete set of 26,116 proof gaps,
Table~\ref{tab:full-benchmark-results} reports Pass@8 scores of
35.39\% for Goedel, 33.27\% for DeepSeek, and 31.99\% for Kimina.
Thus, even the strongest specialized prover leaves 64.61\% of the
local obligations unsolved. ProofGap therefore remains challenging
despite evaluating individual steps rather than complete proofs.

Table~\ref{tab:tactic-baselines} in Appendix~\ref{app:full-passk} examines whether the benchmark can be
solved primarily through standard Lean automation. The strongest
individual tactic, \texttt{aesop}, solves 9.67\% of the gaps, while
the union of all evaluated tactics solves 11.20\%. Moreover,
after removing the 2,925 gaps solved by the tactic union, the learned
provers still achieve Pass@8 scores of 23.57--27.48\% on the
remaining 23,191 gaps. Their performance is therefore not limited to
obligations already handled by the evaluated standard automation.

The topic-level results in Table~\ref{tab:topic-results} show that this
difficulty is structured rather than uniform. All three provers perform
best on single-variable functions, limits, and continuity
(39.10--44.41\%) and worst on parameter integrals
(18.21--20.43\%). The identical topic ordering across the three
provers suggests that these differences reflect stable properties of
the mathematical obligations rather than an isolated weakness of one
model.

\subsubsection{Effect of Proof Granularity (RQ2)}

Under the same eight-candidate budget per target, per-gap Pass@8 is
31.99--35.39\%, whereas complete-theorem Pass@8 is only
2.92--4.28\%. This separation is not caused solely by exercises with
many gaps receiving greater weight. After assigning equal weight to each exercise,
Table~\ref{tab:coverage-by-theorem-outcome} shows that mean
exercise-level gap coverage remains 27.63--31.15\%, whereas
complete-theorem Pass@8 is only 2.92--4.28\%. Thus, step-level
evaluation provides a substantial nonzero signal even when
complete-theorem performance is close to the failure floor.

This comparison is conditional on the context supplied to each gap:
the local target retains facts established earlier in the source
solution, whereas the complete theorem must be proved from its original
premises. The result therefore does not show that gap completion is an
equivalent but easier theorem-proving task. Instead, it separates local
proof construction from the additional demands of organizing and
completing an end-to-end argument. The parameter-integral results make
this distinction particularly visible: the models solve
18.21--20.43\% of the local obligations but at most 1.00\% of the
corresponding complete theorems.

\subsubsection{Cross-Model Comparison (RQ3)}

Table~\ref{tab:proofgap-280-results} shows a clear model-class
separation on ProofGap-280. Every evaluated general-purpose model
outperforms every specialized prover at the gap level. Even GLM-5.2,
the weakest general-purpose model at Pass@8, reaches 65.00\%,
exceeding the strongest specialized prover at 39.64\% by 25.36
percentage points. This result highlights the substantial recent advances in the
formal-reasoning capabilities of general-purpose language models.

The ranking changes substantially on the corresponding parent
exercises. GPT-5.6-sol and Claude-Opus-4.8 retain clear advantages,
and Kimi-K3 is numerically higher than the specialized provers, whereas
DeepSeek-V4-Pro and GLM-5.2 are not. Notably, GPT-5.6-sol and
Claude-Opus-4.8 tie at 87.86\% on the gaps but obtain 21.43\% and
13.93\%, respectively, on the parent exercises. Similar gap-level
performance can therefore coexist with markedly different
end-to-end proof-construction ability.

\subsubsection{Diagnostic Resolution (RQ4)}

Table~\ref{tab:coverage-by-theorem-outcome} shows that gap coverage is
strongly associated with complete-theorem success. Exercises whose
theorem is solved have mean gap coverage of 75.07--79.12\%, compared
with 26.08--29.19\% for theorem-failed exercises. Step-level performance is strongly associated with end-to-end theorem-proving success, while providing a substantially denser diagnostic signal.

At the same time, theorem failure does not imply an absence of local
progress. As shown in Table~\ref{tab:failed-theorem-progress},
72.39--78.03\% of theorem-failed exercises contain at least one
solved gap, 18.79--22.59\% have at least half of their gaps solved,
and 4.46--4.99\% have at least three quarters solved. ProofGap therefore refines a binary theorem failure into a graded account of how many local formal obligations associated with the underlying argument a model can discharge under the evaluated budget.

\Needspace{15\baselineskip}
\section{Conclusion}

We introduced ProofGap, a fine-grained benchmark for step-level formal reasoning based on local proof obligations derived from textbook natural-language solutions. It contains 26,116 gaps from 3,015 mathematical analysis exercises. Experiments show that gaps are substantially more tractable than complete theorems and reveal local progress hidden by theorem-level failure. These findings establish local formal proof completion as a useful interface for fine-grained evaluation and formally grounded diagnostic feedback. The benchmark currently covers one textbook in mathematical analysis; its proof-assistant evaluation is limited to Lean. It evaluates already formalized gaps independently and therefore does not directly measure natural-language autoformalization, invalid-step detection, or end-to-end proof composition. Future work should extend coverage across domains and proof assistants and evaluate complete pipelines that jointly address semantic translation, local proof discharge, and sequential proof composition.

\bibliography{iclr2027_conference}
\bibliographystyle{iclr2027_conference}

\clearpage
\appendix
\begingroup
\hypersetup{hidelinks}
\pdfstringdefDisableCommands{\def\({}\def\){}}
\lstset{style=PGAppendix}
\section{Dataset Construction Details}
\label{app:dataset-construction}

\subsection{Data Sources and Preprocessing}
\label{app:data-sources}

The raw corpus for ProofGap comes from the natural-language problems and
solutions in B. P. Demidovich's \emph{Problems in Mathematical Analysis}. The
book covers topics in mathematical analysis such as functions and limits,
differentiation, integration, series, multivariable analysis, parameter
integrals, and line and surface integrals. Its solutions usually have a clear
step-by-step structure, making them suitable as input to the RNFL pipeline.

Before entering the autoformalization process, the raw text is lightly
preprocessed. First, we remove problems whose understanding necessarily depends
on figures, tables, or external geometric intuition. Second, we split problems
containing multiple mutually independent subquestions into multiple samples, so
that each sample corresponds to a clear problem statement and solution process.
Finally, we normalize the boundaries between problems and solutions, preventing
subsequent problems or residual formatting artifacts from being mixed into the
same sample.

These procedures do not change the mathematical reasoning content of the
original solutions. Instead, they provide structurally clear input for the
subsequent \rnfl{} translation, elaboration, and proof-gap generation.

\FloatBarrier
\Needspace{5\baselineskip}
\subsection{Relaxed NFL Representation and Automatic Translation}
\label{app:relaxed-nfl}

This section explains how natural-language solutions are first represented as
\rnfl{}. \rnfl{} is an intermediate representation between a natural-language
solution and \cnfl{}. It remains close to natural mathematical writing but has
a formal syntax, so it can be read by a parser and processed further. Its role
is not to serve directly as the final verification target, but to provide a
surface language that is easier for large language models to generate while
still preserving proof structure.

\begin{table}[!htbp]
\centering
\caption{Representative natural-language structures preserved by \rnfl{}.}
\label{tab:relaxed-nfl-structures}
\vspace{4pt}
\PGAppendixTableStyle
\begin{tabularx}{\linewidth}{@{}>{\raggedright\arraybackslash}p{.23\linewidth}>{\raggedright\arraybackslash}p{.34\linewidth}>{\raggedright\arraybackslash}X@{}}
\toprule
\textbf{Natural-language structure} & \textbf{\rnfl{} form} & \textbf{Role} \\
\addlinespace[2pt]
\midrule
Known condition &
\code{GIVEN} &
Records problem premises. \\
\addlinespace[2pt]
Goal to prove &
\code{TO PROVE} &
Records the current proof goal. \\
\addlinespace[2pt]
Problem-solving entry &
\code{Solve} &
Records the object, solution set, or range to be found. \\
\addlinespace[2pt]
Intermediate conclusion &
\code{Have P} &
Aligns with a natural-language proof step. \\
\addlinespace[2pt]
Local assumption &
\code{Assume P \{ ... \}} &
Represents conditional reasoning or a nested subproof. \\
\addlinespace[2pt]
Sufficiency transformation &
\code{It suffices to prove P} &
Represents backward reasoning. \\
\addlinespace[2pt]
Local proof of an intermediate proposition &
\code{We prove that P \{ ... \}} &
Generates local proof gaps for $P$ and then makes $P$ available as an
assumption. \\
\addlinespace[2pt]
Candidate answer and checking &

\code{Take t as a witness} \newline
\code{After checking, t satisfies} \newline
\code{the constraints}
 &
Records a candidate object and the verification of its constraints. \\
\addlinespace[2pt]
Method annotation &
\code{[@method M @]} &
Preserves proof-search hints from the source solution. \\
\bottomrule
\end{tabularx}
\end{table}
\FloatBarrier

\rnfl{} resembles natural-language proofs in two main respects. First, it
preserves common reasoning structures in informal proofs, such as stating
intermediate conclusions, reasoning under local assumptions, transforming the
goal, and giving method hints. Second, it permits some terms and propositions
to remain not fully elaborated. For example, surface notation such as $f'$,
$|x|$, and $a_n$, as well as variable scopes introduced by expressions such as
``take arbitrary $x$'' or ``let $x$,'' may first be preserved in \rnfl{} and
then recovered precisely in later stages.

In the ProofGap construction, \rnfl{} preserves at least the structures shown
in Table~\ref{tab:relaxed-nfl-structures}.

A simplified grammar fragment is as follows:

\begin{lstlisting}
<program> ::= <assumptions> Prove: <term> Proof: <proof>
            | <assumptions> Solve: <solving_step>
              Solution: <derivation>

<proof> ::= <derivation> QED

<derivation> ::= <annotated_step>*

<annotated_step> ::= ([@method <method> @])? <step>

<step> ::= Have <term>
         | <action_step>
         | It suffices to prove <term>
         | We prove that <term> { <proof> }
         | Assume <term> { <derivation> }
         | <solving_step> { <derivation> }

<action_step> ::= Set <term> = <term>
                | Take <term> as a witness
                | After checking, <term> satisfies the constraints
                | After checking, <term> does not satisfy the constraints

<solving_step> ::= Find <term> such that <term>
                 | Find all <term> such that <term>
                 | Find the range of <term> such that <term>

\end{lstlisting}

This design makes \rnfl{} more like a structured natural proof than a language
for a particular proof assistant. It reduces the difficulty of the
autoformalization stage while preserving enough proof-structural information
for the subsequent \cnfl{} elaboration.

Based on the surface language above, natural-language problems and solutions
are translated into \rnfl{}. We use prompting-based translation: the prompt
contains the syntax specification of \rnfl{}, target-format requirements, and
several examples from natural language to \rnfl{}, so that the model learns to
preserve proof structure rather than directly generating strict proof assistant
code.

After generation, \rnfl{} is passed through two types of filtering. The first
is parser validation, which ensures that the generated text satisfies the
syntactic constraints of \rnfl{} and can be read by downstream programs. The
second is representation-consistency screening, which filters results that are
clearly inconsistent with the source solution. The screening focuses on the
following points:

\begin{enumerate}
  \item whether the main assumptions in the problem appear in \code{GIVEN} or
  in the corresponding local contexts;
  \item whether the original goal is correctly translated as \code{TO PROVE}
  or \code{Solve};
  \item whether the main proof steps in the natural-language solution are
  represented by \code{Have}, \code{Assume}, \code{We prove that}, or
  \code{It suffices to prove};
  \item whether \rnfl{} introduces key intermediate conclusions that have no
  basis in the source solution;
  \item whether method annotations are broadly consistent with the
  corresponding natural-language steps.
\end{enumerate}

This screening does not claim to prove that the natural-language solution and
\rnfl{} are equivalent in full mathematical semantics. Its purpose is closer
to sanity checking: filtering structural errors, syntax errors, and clearly
misaligned translations. Full verification does not occur at the \rnfl{} stage,
but is realized through the subsequent \cnfl{} elaboration and proof-gap
checking.

\FloatBarrier
\Needspace{5\baselineskip}
\subsection{Core NFL Representation and Verifiable Elaboration}
\label{app:core-nfl}

We first give the surface structure of \cnfl{} and the corresponding
organization of proof states. \cnfl{} is the normalized representation obtained
after elaborating \rnfl{}. Unlike \rnfl{}, which is closer to natural proof
writing, \cnfl{} requires proof steps to eventually be reduced to a small
number of primitive proof constructors. This allows the proof-gap generator to
maintain proof state uniformly along the proof AST and generate proof gaps. We
then describe several finer-grained elaboration procedures used in the
transition from \rnfl{} to \cnfl{}.

Abstractly, \cnfl{} can be viewed as consisting of two syntactic categories:
\[
  \term
    \quad\text{and}\quad
  \method ,
\]
where $\term$ denotes mathematical expressions or propositions in \cnfl{},
and $\method$ denotes proof methods. The special method $?M$ denotes the empty
method.

A proof state consists of the current sequence of assumptions and the current
goal:
\[
  \pstate = \term^{*} \times \term .
\]
If the current assumptions are $\Gamma$ and the goal is $G$, the proof state
is written as
\[
  \Gamma \vdash G .
\]

A \cnfl{} proof consists of a small number of primitive proof constructors:
\[
\begin{aligned}
\mathsf{Proof} ::= {}&
  \mathsf{Forward}(M, P, \pi) \\
&\mid \mathsf{Backward}(M, P, \pi) \\
&\mid \mathsf{Subgoal}(P, \pi_1, \pi_2) \\
&\mid \mathsf{QED}.
\end{aligned}
\]

The meanings of these constructors and their corresponding \rnfl{} structures
are shown in Table~\ref{tab:core-nfl-constructors}.

\begin{table}[!htbp]
\centering
\caption{Primitive \cnfl{} proof constructors and their \rnfl{} origins.}
\label{tab:core-nfl-constructors}
\vspace{4pt}
\PGAppendixTableStyle
\begin{tabularx}{\linewidth}{@{}>{\raggedright\arraybackslash}p{.28\linewidth}>{\raggedright\arraybackslash}p{.29\linewidth}>{\raggedright\arraybackslash}X@{}}
\toprule
\textbf{\cnfl{} constructor} & \textbf{Typical \rnfl{} structure} & \textbf{Meaning} \\
\addlinespace[2pt]
\midrule
\texttt{Forward(M, P, }\(\pi\)\texttt{)} &
\code{[@method M] Have P} &
Derives an intermediate proposition $P$ from the current proof state using
method $M$, and adds $P$ to the subsequent assumptions. \\
\addlinespace[2pt]
\texttt{Backward(M, P, }\(\pi\)\texttt{)} &
\code{[@method M] It suffices to prove P} &
Reduces the current goal to the subgoal $P$ using method $M$, then continues
by proving $P$. \\
\addlinespace[2pt]
\texttt{Subgoal(P, }\(\pi_1\)\texttt{, }\(\pi_2\)\texttt{)} &
\code{We prove that P \{ ... \}} &
First proves the intermediate subgoal $P$ with $\pi_1$, then uses $P$ as a
known fact in $\pi_2$. \\
\addlinespace[2pt]
\code{QED} &
\code{QED} &
Ends the current proof, indicating that the current goal follows from the
current assumptions. \\
\bottomrule
\end{tabularx}
\end{table}
Thus, a \cnfl{} program can be regarded as
\[
  \mathsf{Program} = \pstate \times \mathsf{Proof}.
\]

The \cnfl{} column in Figure~1 of the main paper displays a readable
surface form of this structure in data construction: it has already made
variable scopes, local conditions, and context-dependent notation explicit,
and can be consumed by the proof-gap generator. Internally, the system does not
treat it as an ordinary text sequence, but parses it into a proof AST
consisting of the proof-state structure and primitive proof constructors
described above.

The definitions above explain how the main proof steps in \rnfl{} are reduced
to the core constructors required by \cnfl{}. In actual construction,
finer-grained elaboration is also needed, such as recovering implicit variable
scopes, eliminating ambiguities in context-dependent notation, splitting
compound propositions, and normalizing derivative and integral notation in
mathematical analysis. These processes are mainly performed by deterministic
transformations. In the few places where the intent of the original
natural-language proof must be judged, the system uses constrained LLM
heuristics and requires the output to pass programmatic checks. We give several
representative types of elaboration below.

\subsubsection{Notation Disambiguation}

A common source of ambiguity in natural mathematical writing is notation
overloading: the same symbol may denote different mathematical objects. A
simple example is $|x|$. Depending on the meaning of $x$, it may denote either
the absolute value of a real number or the cardinality of a set. Therefore, the
surface notation $|x|$ in \rnfl{} cannot by itself be automatically fixed as a
particular \cnfl{} term.

To perform notation disambiguation, the system preserves overloaded notation
in the term language of \rnfl{} while introducing unambiguous forms and
auxiliary annotations:

\begin{lstlisting}
<term> ::= ...
         | |<term>|
         | IsReal(<term>)
         | IsSet(<term>)
         | Abs(<term>)
         | Card(<term>)
\end{lstlisting}

Here, \code{|.|} is overloaded surface notation, while \code{Abs(.)} and
\code{Card(.)} are unambiguous forms that can be deterministically translated
to \cnfl{}. The annotations \code{IsReal(.)} and \code{IsSet(.)} record
whether a term is regarded as a real number or as a set. They are not a static
type system for \rnfl{}, but annotations used to guide the interpretation of
overloaded notation.

The elaborator performs static analysis over \rnfl{} and maintains an
environment recording these annotations. The annotations may come from two
sources. The first is information explicitly preserved by the model during the
initial autoformalization stage; for example, ``for every real number $t$''
introduces \code{IsReal(t)}. The second is constraint propagation performed by
the system according to operator signatures; for example, when an operator
requires a set argument, an \code{IsSet} annotation is generated for the
corresponding object. When the environment determines a unique interpretation,
the system directly replaces the overloaded notation:
\[
  |t| \mapsto \mathsf{Abs}(t)
  \quad\text{if } \mathsf{IsReal}(t),
\]
\[
  |t| \mapsto \mathsf{Card}(t)
  \quad\text{if } \mathsf{IsSet}(t).
\]

If multiple candidate interpretations remain after static constraints are
applied, the system gives the model only the local proof context and a finite
candidate set, asks it to select the intended interpretation, and then checks
whether the choice is consistent with the current constraint environment.

The same disambiguation mechanism is also used for other context-dependent
notation in mathematical analysis. For example, $f'$ must be interpreted as a
function derivative or some other derived object; $a_n$ may denote the $n$-th
term of a sequence, or a partial derivative of a function or multivariate
object with respect to $n$; and $f^{-1}$ may denote an inverse function, a
matrix inverse, or an ordinary power of $-1$. The system selects the
interpretation using the local object category, operator signatures, and
contextual constraints, and uses constrained candidate selection when needed.

\subsubsection{Splitting Chained Equalities and Compound Propositions}

Natural-language proofs often compress several atomic relations into a single
compound proposition, such as a chained inequality, a chained equality, or a
conjunction. \rnfl{} may preserve this writing style.However, when generating proof gaps, a compound proposition that is too large mixes multiple local inferences into the same goal, making it difficult to
obtain fine-grained step-level obligations. Therefore, elaboration normalizes
such compound structures into smaller verification units:

\begin{align*}
\mathsf{Have}\ a < b < c
  &\rightsquigarrow
    \mathsf{Have}\ a < b,\quad
    \mathsf{Have}\ b < c,\quad
    \mathsf{Have}\ a < c, \\
\mathsf{Have}\ a = b = c
  &\rightsquigarrow
    \mathsf{Have}\ a = b,\quad
    \mathsf{Have}\ b = c,\quad
    \mathsf{Have}\ a = c, \\
\mathsf{Have}\ P \land Q
  &\rightsquigarrow
    \mathsf{Have}\ P,\quad
    \mathsf{Have}\ Q.
\end{align*}

This transformation does not perform proof search, nor does it try to prove
the individual subpropositions. It only unfolds surface-level compound
propositions into more explicit \cnfl{} steps. After expansion, each
subproposition can generate an independent proof gap under the current proof
state. This preserves the locality of the original proof step while making
error localization more precise.

\subsubsection{Recovering Implicit Variable Scope}

\rnfl{} permits temporarily unbound variables to appear in proof steps. For
example, \code{Have P(x)} only states that the current step claims $P(x)$, but
it does not yet specify whether $x$ is an arbitrary real number, a fixed
object, or a variable satisfying some local conditions. \cnfl{} needs such
implicit scopes to be made explicit; otherwise, the generated proof gap lacks
complete context.

For this purpose, the system first marks free variables in propositions with
\code{[@free ... @]}, and then marks the implicit scope these variables should
inhabit with \code{[@scope ... @]}. The scope annotation is then eliminated and
converted into explicit quantifiers and conditions in \cnfl{}. For example:

\Needspace{14\baselineskip}
\begin{lstlisting}
Free-variable annotation:
[@free x @] Have P(x)

Scope annotation:
[@scope [forall x] [x in Real] @] {
  Have P(x)
}

After scope elimination:
Have forall (x), x in Real =>
  P(x)
\end{lstlisting}

Here, \code{[@free x @]} indicates that $x$ has not yet been explicitly bound
by the current proposition, while \code{[@scope [forall x] [x in Real] @]}
indicates that the step should be interpreted as holding ``for every real
number $x$.'' The model output for this step must pass a deterministic checker:
all annotated free variables must fall within matching scopes, and apart from
adding and eliminating scope annotations, no other part of the proof may be
modified.

\subsubsection{Canonicalizing Derivatives and Indefinite Integrals}

Because ProofGap targets mathematical analysis, it also needs a unified
internal canonical form for derivatives and indefinite integrals. The goal here
is not to execute differentiation or integration, but to translate surface
notation in natural writing into semantically explicit \cnfl{} terms without
free variables.

Derivatives are uniformly represented as
\[
  \mathsf{FunDeri}(f,g,n),
\]
denoting the $n$-th derivative of function $f$ with respect to function $g$.
In this way, ordinary derivatives, partial derivatives, higher-order
derivatives, and derivatives with respect to a composite function can all be
placed in the same canonical form. For example, the derivative of $\sin(x)$
with respect to $\cos(x)$,
\[
  \frac{d(\sin x)}{d(\cos x)},
\]
is written in \cnfl{} as
\[
  \mathsf{FunDeri}(\lambda x.\sin x,\lambda x.\cos x,1).
\]

As another example, for the second-order partial derivative of a binary
function $f(x,y)$ with respect to $y$, the natural notation
\[
  \frac{\partial^2 f(x,y)}{\partial y^2}
\]
can first be represented as
\[
  \mathsf{FunDeri}(f,\lambda x,y.\,y,2).
\]
Since $\lambda x,y.\,y$ selects only the second formal parameter, this
representation can also be written in a more readable form:
\[
  \mathsf{FunDeri}(\lambda x,y.\,f(x,y),2,2).
\]
These two \cnfl{} terms have the same semantics; the latter only displays
``differentiate with respect to the second formal parameter'' more concisely.

The canonicalization of indefinite integrals is based on another consideration:
an indefinite integral denotes a family of antiderivatives, not a single
function value. For example, the natural notation
\[
  \int 2x\,dx = x^2 + C
\]
is represented in \cnfl{} as an equality between two sets of functions:
\[
  \{\,F \mid \forall x,\ \mathsf{FunDeri}(F,1,1)(x)=2x\,\}
  =
  \{\,F \mid \exists C \in \mathbb{R},\
    \forall x,\ F(x)=x^2+C\,\}.
\]
This treatment is necessary: the result of $\int 2x\,dx$ is not a fixed
function, but the set of all functions satisfying $F'(x)=2x$; the right-hand
side $x^2+C$ also denotes a family of functions parameterized by an arbitrary
real constant $C$. Thus, in \cnfl{}, an equality of indefinite integrals is
naturally represented as an equality between two sets of functions.

\FloatBarrier
\Needspace{5\baselineskip}
\subsection{Proof-Gap Generation and Checking}
\label{app:gap-generation}

After verifiable elaboration, the semantics of \cnfl{} is given by proof-gap
generation. That is, we do not interpret a \cnfl{} program directly as a
complete proof object; instead, we interpret it as a process that generates
verification conditions. If all generated proof gaps are successfully proved,
then the \cnfl{} program is regarded as verified.

A proof gap consists of a proof state and a method:
\[
  \gap = \pstate \times \method .
\]
If the proof state is $\Gamma \vdash G$ and the method annotation is $M$, the
gap is written as
\[
  \Gamma \vdash_M G .
\]
It means: under assumptions $\Gamma$, prove the goal $G$. The method $M$ is a
hint preserved from the natural-language solution and does not change the
formal validity of the proof obligation.

The proof-gap generator $\pgg$ is defined recursively according to the
structure of the \cnfl{} proof. Let $\Gamma;P$ denote appending proposition
$P$ to the end of the assumption sequence. The generation rules are:
\begin{align*}
\pgg(\Gamma \vdash G,\ \mathsf{Forward}(M,P,\pi))
  &=
  \{\,\Gamma \vdash_M P\,\}
  \cup \pgg(\Gamma;P \vdash G,\ \pi),
\\
\pgg(\Gamma \vdash G,\ \mathsf{Backward}(M,P,\pi))
  &=
  \{\,\Gamma;P \vdash_M G\,\}
  \cup \pgg(\Gamma \vdash P,\ \pi),
\\
\pgg(\Gamma \vdash G,\ \mathsf{Subgoal}(P,\pi_1,\pi_2))
  &=
  \pgg(\Gamma \vdash P,\ \pi_1)
  \cup \pgg(\Gamma;P \vdash G,\ \pi_2),
\\
\pgg(\Gamma \vdash G,\ \mathsf{QED})
  &=
  \{\,\Gamma \vdash_{?M} G\,\}.
\end{align*}

This definition explains why ProofGap is a local benchmark: every gap comes
from a proof constructor in \cnfl{} rather than from an arbitrary subgoal
extracted from a complete theorem. An intermediate step in the natural-language
solution corresponds to a proof-state transition in \cnfl{} and is then
expanded into a local proof obligation.

Furthermore, we define
\[
  \checkgap : \gap \to \{\mathsf{false},\mathsf{true}\},
\]
and
\[
  \checkprogram(\sigma,\pi)
  =
  \bigwedge_{H \in \pgg(\sigma,\pi)} \checkgap(H).
\]

In this benchmark, $\checkgap$ can be implemented through two compatible
interfaces: the proof-checking framework over the lightweight ProofGap
representation, and the corresponding Lean theorem skeleton. During
evaluation, a model must generate, for each proof gap, a formal derivation
accepted by the selected interface.

\Needspace{18\baselineskip}
\FloatBarrier
\Needspace{5\baselineskip}
\subsection{ProofGap Data Fields and Lean Skeleton Export}
\label{app:data-fields}

A released ProofGap instance expands the sequent form above into data fields:

\Needspace{9\baselineskip}
\begin{lstlisting}
ASSUM:
  Gamma

GOAL:
  G

METHOD:
  M
\end{lstlisting}

The \code{METHOD} field is optional. It records strategy hints from the source
natural-language solution, such as ``differentiate both sides,'' ``by
monotonicity,'' or ``compare coefficients.'' It is withheld in all reported
evaluations and released for future analyses of strategy-conditioned proving.

Since each ProofGap sample already contains explicit local assumptions and a
goal, it can be exported relatively directly as a Lean theorem skeleton. The
Lean version and the lightweight ProofGap representation are two compatible
interfaces for the same local proof obligation: the former supports manual
inspection, reuse, and connection to the ecosystem of existing interactive
theorem provers, while the latter provides a compact benchmark-native
representation.

For example, the following ProofGap:

\Needspace{12\baselineskip}
\begin{lstlisting}
ASSUM:
DiffableFunc(f)
forall x, EvenFunc(f) => f(x)=f(-x)

GOAL:
forall x, EvenFunc(f)
  => FunDeri(f,1,1)(x)= -FunDeri(f,1,1)(-x)

METHOD:
differentiate both sides
\end{lstlisting}

can be exported as the following Lean theorem skeleton:

\Needspace{12\baselineskip}
\begin{lstlisting}
theorem gap2
    (f : Real -> Real)
    (h1 : Differentiable Real f)
    (h2 : forall x : Real, Function.Even f
      -> f x = f (-x)) :
    forall x : Real, Function.Even f ->
      deriv f x = -deriv f (-x) := by
  sorry
\end{lstlisting}

This export is mainly structural: ProofGap has already split the local proof
obligation into \code{ASSUM} and \code{GOAL}, so exporting a Lean skeleton only
requires translating each assumption into a theorem parameter and placing the
goal in the conclusion position of the theorem. In the example above,
\code{DiffableFunc(f)} is exported as the differentiability premise \code{h1},
the local fact \code{forall x, EvenFunc(f) => f(x)=f(-x)} is exported as
premise \code{h2}, and \code{GOAL} becomes the theorem conclusion. Thus, the
Lean form and the lightweight ProofGap form are not different tasks, but two
representations and checking interfaces for the same proof-gap instance.

\FloatBarrier
\Needspace{5\baselineskip}
\subsection{Quality Control and Traceability}
\label{app:quality-control}

Before a sample enters the final benchmark, it must pass several layers of
programmatic checks:

\begin{enumerate}
  \item the boundaries of the natural-language problem and solution are clear,
  and the sample does not depend on unrecoverable graphical information;
  \item \rnfl{} can be parsed by the parser;
  \item \rnfl{} has basic proof-structure consistency with the source solution;
  \item elaboration from \rnfl{} to \cnfl{} completes successfully;
  \item free-variable annotations and scope annotations have been eliminated,
  and the binding-completeness check passes;
  \item \cnfl{} can generate nonempty, parseable \code{ASSUM} and \code{GOAL}
  fields.
\end{enumerate}

These checks cannot guarantee that the autoformalized result is absolutely
correct in full mathematical semantics, but they keep the construction process
controlled and verifiable. \rnfl{} serves as an intermediate representation
aligned with natural language, \cnfl{} provides a semantically explicit
proof-state structure, and proof gaps turn the correctness of a complete proof
into a set of locally checkable verification conditions. It is important to
emphasize that the entire construction process operates on syntax trees
generated by the parser: elaboration transforms the \rnfl{} AST, while the
proof-gap generator maintains proof state and generates proof obligations along
the \cnfl{} proof AST. Therefore, each proof gap preserves its generation
position in the proof AST and can be mapped back to the corresponding local
proof step in the source solution.

% Detailed appendix for the ProofGap lightweight checking framework.

\FloatBarrier
\Needspace{5\baselineskip}
\section{Verification Framework and Agent Baseline}
\label{app:framework-details}

This appendix expands the compact framework description in the main paper.  It
records the complete DSL command inventory, the checker's operational trust
boundary, explicit solver coverage, theorem-instantiation protocol, and the
controller used by the verifier-guided baseline.  The purpose is to make
successful certificates replayable and the system's failure modes explicit
without enlarging the main-paper presentation.

\subsection{DSL Command Inventory}
\label{app:dsl-reference}

The DSL contains 18 commands, grouped in Table~\ref{tab:dsl-commands}.  Its
granularity lies between an end-to-end formal proof and an unconstrained
natural-language step: one instruction performs a recognizable reasoning
operation, while the checker validates its exact effect.  In particular,
\texttt{autosolve} is itself a DSL command rather than an external preprocessing
step.  Automation can receive either every local hypothesis or an explicit
model-selected subset; explicit selection records the dependencies of a step
and limits accidental reliance on irrelevant context.

\begin{table}[!htbp]
\centering
\caption{The 18 DSL commands, organized by their role in proof-gap
completion.}
\label{tab:dsl-commands}
\vspace{4pt}
\PGAppendixTableStyle
\begin{tabularx}{\linewidth}{@{}>{\raggedright\arraybackslash}p{.19\linewidth}>{\raggedright\arraybackslash}p{.37\linewidth}>{\raggedright\arraybackslash}X@{}}
\toprule
\textbf{Command family} & \textbf{Commands} & \textbf{Effect on the proof state} \\
\addlinespace[2pt]
\midrule
Discharge &
\texttt{autosolve}, \texttt{apply} &
Close the goal with a solver or reduce it to the unmatched premises of an
applicable hypothesis. \\
\addlinespace[2pt]
Fact management &
\texttt{get\_prop}, \texttt{use\_condition}, \texttt{assert},
\texttt{destruct\_and} &
Instantiate a hypothesis or theorem, eliminate implication premises, establish
a checked local claim, or decompose a conjunction. \\
\addlinespace[2pt]
Quantifiers and implications &
\texttt{exists}, \texttt{get\_exists}, \texttt{get\_forall},
\texttt{get\_condition} &
Provide witnesses, open quantified hypotheses or goals, and move an antecedent
from a goal into the local context. \\
\addlinespace[2pt]
Rewriting and normalization &
\texttt{autoreplace}, \texttt{not\_normalize}, \texttt{rewrite},
\texttt{rewrite\_in limit}, \texttt{rewrite\_in sum},
\texttt{rewrite\_in setdesc} &
Perform checked transformations, including binder-aware rewriting under
limits, sums, and set comprehensions. \\
\addlinespace[2pt]
Branching &
\texttt{case\_analysis}, \texttt{or\_intro} &
Split on a disjunctive hypothesis or select a side of a disjunctive goal. \\
\bottomrule
\end{tabularx}
\end{table}
\FloatBarrier

\paragraph{Surface syntax.}
Programs are line-oriented; formulas are enclosed in braces, hypotheses carry
explicit indices, and a hash mark starts a comment.  The following schematic
forms summarize the complete command surface.  Here \texttt{source} and
\texttt{result} are a hypothesis or the goal, \texttt{dir} is a rewrite
direction, and ellipses denote repeated binders or hypotheses rather than
unchecked text.

\begin{lstlisting}[numbers=none]
autoreplace <scope>; from <term> to <term>; from <source> to <result>
not_normalize from (GOAL: {...}) to (GOAL: {...})
autosolve all_hypothesis
autosolve (hypothesis <i>: {...}) ...
get_prop (<hypothesis-or-theorem>) (<x> := {...}) ... as (hypothesis <j>: {...})
exists (<x> := {...}) ...
destruct_and (hypothesis <i>: {...}) as (hypothesis <j>: {...}) ...
rewrite (<dir>, <equation-or-iff>) (hypothesis <i>: {...}); from <source> to <result>
rewrite_in limit (<dir>, equation) (hypothesis <i>: {...}); from <source> to <result>
rewrite_in sum (<dir>, equation) (hypothesis <i>: {...}); from <source> to <result>
rewrite_in setdesc (<dir>, equation) (hypothesis <i>: {...}) (<x> := {...}); from <source> to <result>
apply (hypothesis <i>: {...})
use_condition (<condition>) ... in (hypothesis <i>: {...}) as (hypothesis <j>: {...})
get_exists (hypothesis <i>: {...}) (<x> := {...}) ...
get_forall (GOAL: {...}) (<x> := {...}) ...
get_condition (hypothesis <j>: {...}) ... in (GOAL: {...})
assert (hypothesis <j>: {...}) [by proof { ... }]
case_analysis (hypothesis <i>: {...})
or_intro <left-or-right>
\end{lstlisting}

\paragraph{Command-level semantics.}
The checker assigns the following precise state transition to each command.

\begin{enumerate}
  \item \textbf{\texttt{autoreplace}.} Recompute a supported local
  transformation in a hypothesis or the goal, using no hypotheses, all
  hypotheses, or an explicit subset.  The command states the replaced term,
  its replacement, the complete source formula, and the claimed result; an
  arbitrary user-written substitution is not trusted.

  \item \textbf{\texttt{not\_normalize}.} Replace only the goal, and only when
  the old and new forms have $\alpha$-equivalent fully unfolded negation normal
  forms.  The command rejects a no-op and invokes neither general automation
  nor linear arithmetic.

  \item \textbf{\texttt{autosolve}.} Close the current goal only when the
  deterministic solver portfolio certifies it from all hypotheses or the
  explicitly selected subset.  Thus \texttt{autosolve} is an ordinary DSL
  instruction whose success can be replayed.

  \item \textbf{\texttt{get\_prop}.} Instantiate outer universal binders of a
  hypothesis or a named library theorem and add the checked instance as a new
  hypothesis.  The checker requires enough outer binders, performs
  capture-avoiding simultaneous substitution and beta reduction, and compares
  the supplied result modulo $\alpha$-equivalence.

  \item \textbf{\texttt{exists}.} Supply witnesses for one or more consecutive
  outer existential binders of the goal.  The stated binder names must match
  those binders, and the residual goal is recomputed by the checker.

  \item \textbf{\texttt{destruct\_and}.} Recursively flatten a conjunctive
  hypothesis.  The number and order of explicitly named output hypotheses must
  exactly match all flattened conjuncts.

  \item \textbf{\texttt{rewrite}.} Apply a directed equation or biconditional
  from a hypothesis to a goal or hypothesis.  It checks the complete before
  and after formulas and does not cross a protected binder.  A rewrite rule
  still wrapped in an outer quantifier must first be instantiated.

  \item \textbf{\texttt{rewrite\_in limit}.} Rewrite one selected limit body
  using a hypothesis with exactly one outer universal binder and an equation
  conclusion.  A conditional rule must hold eventually in the relevant
  punctured, one-sided, or infinite neighborhood.

  \item \textbf{\texttt{rewrite\_in sum}.} Rewrite one selected summand with
  the same single-binder equation form, after proving the rule condition for
  every index in the entire finite or supported infinite summation range.

  \item \textbf{\texttt{rewrite\_in setdesc}.} Rewrite within one selected set
  comprehension using an explicit map from the rule binder to the set binder.
  The side condition may use unaffected local set conditions but not the very
  condition being rewritten.

  \item \textbf{\texttt{apply}.} Close a matching goal or reduce it to the
  unmatched prefix of premises of an implication chain.  Any resulting
  premises remain explicit proof obligations.

  \item \textbf{\texttt{use\_condition}.} Discharge an ordered prefix of an
  implication hypothesis and add the residual implication or conclusion as a
  new hypothesis.  Each premise is supplied by a matching hypothesis or by an
  \texttt{obvious} fact that is itself checked.

  \item \textbf{\texttt{get\_exists}.} Open consecutive outer existentials in
  a hypothesis using fresh names and update that hypothesis in place.

  \item \textbf{\texttt{get\_forall}.} Open consecutive outer universal
  binders of the goal using fresh names and replace the goal with the checked
  body.

  \item \textbf{\texttt{get\_condition}.} Move an ordered prefix of goal
  antecedents into the local context under explicitly assigned fresh
  hypothesis indices.

  \item \textbf{\texttt{assert}.} Add an intermediate proposition only after
  it is discharged either by automation or by a nonempty nested DSL proof.
  The nested proof receives a copy of the outer assumptions; hypotheses made
  inside it cannot leak into the outer state.

  \item \textbf{\texttt{case\_analysis}.} Split a disjunctive hypothesis into
  one active proof gap for each recursively exposed disjunct.  Acceptance
  requires every resulting branch to close.

  \item \textbf{\texttt{or\_intro}.} Select the left or right side of a binary
  disjunctive goal and replace the goal by that selected proposition.
\end{enumerate}

\subsection{Checker Semantics and Trust Boundary}
\label{app:checker-semantics}

The checker implements the operational semantics of these commands as
deterministic proof-state transitions.  Commands that transform a formula state
both the source and intended result; the source must match the current state,
and the checker recomputes or validates the claimed change.  Quantifier
instantiation uses capture-avoiding substitution and compares expressions
modulo $\alpha$-equivalence.  Fresh-variable and hypothesis-index constraints
make the execution trace unambiguous.  The \texttt{assert} command may contain
a nested DSL program, but facts created inside that local proof do not escape;
only its checked conclusion is added to the outer state.  Acceptance requires
every active goal to close, so a program cannot leave a case-analysis branch or
local assertion unfinished.

Rewriting beneath implicit binders receives additional checks.  Ordinary
\texttt{rewrite} is binder protecting; the three \texttt{rewrite\_in}
commands are explicit, scoped exceptions.  Their rule must be a hypothesis
with exactly one outer universal binder and an equation conclusion.  Before
substitution, conflicting binders are $\alpha$-renamed, and a candidate rule
instance that depends on a binder introduced locally by the rule is rejected.
Every unchanged sibling, binder, limit point, summation bound, and set
condition must remain $\alpha$-equivalent, so algebraic normalization cannot
be hidden as an additional rewrite.  Equality must hold eventually in the
relevant neighborhood for a limit, at every index---including initial
indices---for a sum, or under the unaffected local conditions of a set
comprehension.  Excluding the condition currently being rewritten prevents a
circular, self-justifying set rewrite.  These checks are particularly
important in analysis, where an otherwise plausible rewrite may capture a
variable or omit a domain condition.

The parser, checker, and solver have distinct failure contracts.  A parse error
produces no proof state; an illegal transition leaves the current state
unaccepted; and a solver that cannot establish its obligation produces no
certificate.  In particular, unsupported solver input is reported as
\emph{unknown}.  Execution reports the first
failed command together with the remaining local context and goal.  Thus
syntax failures, unsupported automation, unfinished branches, and
mathematically invalid transitions remain separable in downstream analysis.

\subsection{Solver Portfolio and Capability Boundaries}
\label{app:solver-boundaries}

The DSL command \texttt{autosolve} invokes a portfolio of specialized,
deterministic solvers over the selected hypotheses.  Selected backends may also
be requested explicitly; for example, the interface exposes
\texttt{LimitCalcSolver}.  An unsupported expression is reported as
\emph{unknown}; an unresolved side condition or exhausted bounded search
likewise produces no certificate.  This conservative contract permits
composition without turning an unsupported expression into a proof.

\begin{table}[!htbp]
\centering
\caption{Major solver families and their supported fragments; the table does
not claim completeness for the corresponding mathematical theories.}
\label{tab:solver-capabilities}
\vspace{4pt}
\PGAppendixTableStyle
\begin{tabularx}{\linewidth}{@{}>{\raggedright\arraybackslash}p{.20\linewidth}>{\raggedright\arraybackslash}p{.37\linewidth}>{\raggedright\arraybackslash}X@{}}
\toprule
\textbf{Capability family} & \textbf{Representative solvers} & \textbf{Supported reasoning} \\
\addlinespace[2pt]
\midrule
Logical and structural &
\texttt{EqbSolver}, \texttt{Reflexivity}, \texttt{NotUnfoldSolver},
\texttt{FunctionApplySolver} &
Hypothesis matching modulo $\alpha$-equivalence, reflexive relations, negation
normalization, beta reduction, definition unfolding, and deterministic
piecewise-function evaluation. \\
\addlinespace[2pt]
Domains and constants &
\texttt{ConstTypeSolver}, \texttt{DomainTypeSolver}, shared exact-numeric
analysis &
Membership in standard number sets, closure under contextual type facts, and
exact evaluation of closed constants subject to definedness checks. \\
\addlinespace[2pt]
Algebra and inequalities &
\texttt{PolyRatSolver}, \texttt{RootFracSolver}, \texttt{RootPowerSolver},
\texttt{SgnPureSolver}/\allowbreak\texttt{SgnSolver}, \texttt{IntervalSolver},
\texttt{SqueezeSolver}, \texttt{SmtLraSolver} &
Polynomial and rational identities, root and fraction normalization, sign
propagation, linear arithmetic, interval reasoning, and bound propagation. \\
\addlinespace[2pt]
Calculus and analysis &
\texttt{DerivSolver}, \texttt{ContinuousFuncSolver},
\texttt{IntegrableFuncSolver}, \texttt{IntegralSolver},
\texttt{EvalBarSolver}, \texttt{LimitCalcSolver} &
Symbolic derivatives, elementary continuity, integrability on finite closed
intervals, antiderivative verification, endpoint evaluation, and conservative
finite-point or positive-infinity limits. \\
\bottomrule
\end{tabularx}
\end{table}
\FloatBarrier

\paragraph{Logical reduction and domain analysis.}
\texttt{EqbSolver} matches a goal against an assumption modulo
$\alpha$-equivalence, while the separate reflexivity backend handles
relations such as equality and subset inclusion.  \texttt{FunctionApplySolver}
performs capture-avoiding simultaneous beta reduction, unfolds admissible
nonrecursive closed definitions, and selects a piecewise point value only when
the selected guard is proved and every preceding guard is disproved.
\texttt{ConstTypeSolver} is restricted to number-set membership of closed
constant terms; concrete primality is checked only up to $10^6$.
\texttt{DomainTypeSolver} instead propagates contextual membership through
compound expressions while retaining nonzero, root, and power conditions.

\paragraph{Algebra and inequalities.}
Polynomial and rational normalization, root--fraction normalization, and
root--power conversion are separated into specialized backends.  Sign solvers
propagate positive, negative, zero, and nonzero information through supported
expression structure.  \texttt{IntervalSolver} performs interval-constraint
propagation and distinguishes defined, undefined, and unknown closed
expressions.  \texttt{SqueezeSolver} uses explicit comparison and sign facts
in a bounded local search; it neither invokes another solver to manufacture a
side condition nor performs unrestricted algebraic expansion.  Linear real
arithmetic is available through a separate backend rather than being silently
assumed by every transformation.

\paragraph{Calculus and analysis.}
The calculus backends certify restricted, explicit fragments rather than
invoking a general computer-algebra oracle.  \texttt{DerivSolver} evaluates
\texttt{FunDeri} expressions from lambdas or admissible nonrecursive
definitions and supports higher and mixed partials; it does not invent
derivatives for unknown symbolic functions.  Conditional derivative and
pointwise-continuity definitions must hold on a real open neighborhood of the
evaluation point, not merely at that point.  \texttt{ContinuousFuncSolver}
handles resolved unary lambdas, arithmetic closure, and supported elementary
compositions subject to their natural domains.  \texttt{IntegrableFuncSolver}
automatically proves integrability only on finite closed intervals, using
continuity as a sufficient condition.  \texttt{IntegralSolver} verifies a
normalized antiderivative family by differentiation and requires an explicit,
unconstrained real integration constant; it does not synthesize an indefinite
integral.  \texttt{EvalBarSolver} checks finite endpoint evaluation of an
explicit lambda after validating division, powers, logarithms, roots, and
tangent at both endpoints.  \texttt{LimitCalcSolver} supports
continuity-based finite substitution and conservative positive-infinity
rational or growth-order cases, but not negative-infinity and general
one-sided limits, L'Hopital's rule, or general asymptotic equivalence.

\paragraph{Definedness and exactness.}
Solver obligations retain their domain information.  Division requires a
nonzero denominator, logarithms require admissible arguments and bases, even
roots require nonnegative radicands, and a piecewise branch is selected only
under the guarded conditions above.  Substitution is binder-aware, preventing
a fact about a free variable from being reused for a shadowing bound variable.
Rational expressions are evaluated exactly when possible, while constants
such as $e$ and $\pi$ use certified rational intervals for comparisons rather
than ordinary floating-point approximation.  Unknown precision or an
unsupported construct is never silently converted into a proof.

\subsection{Theorem Library}
\label{app:theorem-library}

Many proof gaps require named mathematical knowledge beyond computation and
decision procedures.  We therefore provide a theorem library in the same
formal language as the proof states.  The packaged solutions use the included
theorem-library snapshot, which contains 514
consecutively indexed DSL entries.  They cover sets, binary relations,
functions, cardinality, countability, and induction; limits, continuity,
differentiation, integration, sequences, and series; algebraic, trigonometric,
exponential, logarithmic, hyperbolic, inverse-trigonometric, and complex
identities; and reusable definitions for intervals, extrema, function and
sequence properties, and multivariable differential operators.  The snapshot
contains high-level results such as the mean-value and squeeze theorems,
definition-expansion and bridge lemmas for short local steps, and
alpha-compatible variants of frequently used declarations.

Table~\ref{tab:library-coverage} gives a descriptive organization by index
range.  These ranges summarize the bundled sequence; they are not additional
types trusted by the checker.

\begin{table}[!htbp]
\centering
\caption{Descriptive coverage of the bundled 514-entry theorem-library
snapshot.}
\label{tab:library-coverage}
\vspace{4pt}
\PGAppendixTableStyle
\begin{tabularx}{\linewidth}{@{}rr>{\raggedright\arraybackslash}X@{}}
\toprule
\textbf{Indices} & \textbf{Count} & \textbf{Descriptive coverage} \\
\addlinespace[2pt]
\midrule
1--75 & 75 & Sets and binary relations \\
\addlinespace[2pt]
76--109 & 34 & Functions, cardinality, and natural numbers \\
\addlinespace[2pt]
110--147 & 38 & Core analysis results and proof rules \\
\addlinespace[2pt]
148--241 & 94 & Algebraic, trigonometric, exponential, and complex identities \\
\addlinespace[2pt]
242--277 & 36 & Auxiliary mathematical operators and definitions \\
\addlinespace[2pt]
278--352 & 75 & Function properties, continuity, differentiability, sequences, and series \\
\addlinespace[2pt]
353--494 & 142 & Executable bridge lemmas for analysis reasoning \\
\addlinespace[2pt]
495--514 & 20 & Alpha-variable compatibility entries \\
\bottomrule
\end{tabularx}
\end{table}

Each entry pairs a stable numeric identifier and a human-readable title with a
formal DSL statement.  Titles are not assumed to be globally unique, so the
numbered identifier is the primary reference.  Library entries and proof gaps
pass through the same semantic parser, so intervals, indexed sums, and set
comprehensions receive the same AST-level interpretation.  To use an entry, a
DSL program invokes \texttt{get\_prop}, quotes the entry, optionally supplies
explicit substitutions for a subset of its outer universal binders, and states
the resulting hypothesis.  The checker resolves the numbered entry, performs
capture-avoiding instantiation and beta reduction, and accepts the hypothesis
only when it is $\alpha$-equivalent to the recomputed instance.  Its premises
must then be discharged explicitly.  A model therefore cannot cite a
nonexistent theorem or silently discard its side conditions, and missing
library knowledge remains an observable failure mode.

This 514-entry theorem snapshot is included in the final artifact and used to
replay its packaged solutions.  Among the 7,727 accepted programs, 569 cite at
least one numbered library entry; they contain 1,053 references spanning 180
distinct theorem identifiers.  The library is thus used selectively alongside
solver- and context-driven certificates.

\subsection{Unified Verifier-Guided Agent Workflow}
\label{app:agent-workflow}

The verifier-guided baseline integrates complementary
mechanisms explored by four workflow implementations behind one task and
checker interface; it does not assume that every implementation contains every
mechanism.  The language model proposes and revises DSL programs, whereas a
deterministic controller atomically claims gaps, pins the checker and
theorem-library versions, assigns budgets, and executes each candidate in an
isolated work area.  Attempts retain their DSL, first failure, residual state,
and terminal decision, enabling replay and recovery.
Figure~\ref{fig:agent-baseline} summarizes the workflow.

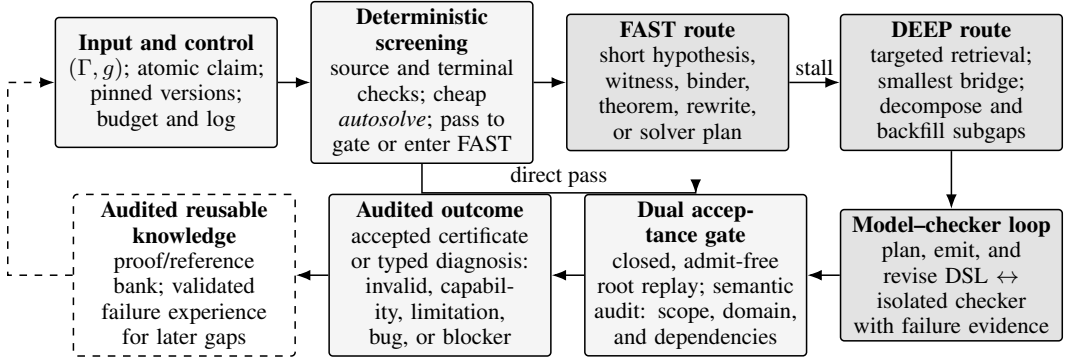
\begin{figure}[!htbp]
\centering
\resizebox{\linewidth}{!}{%
\begin{tikzpicture}[
  >=Latex,
  every node/.style={font=\small},
  stage/.style={draw, line width=0.6pt, rounded corners=1.5pt, align=center,
    text width=.206\textwidth, minimum height=18mm,
    inner xsep=2.8pt, inner ysep=2.8pt},
  deterministic/.style={stage, fill=black!4},
  model/.style={stage, fill=black!11},
  store/.style={stage, dashed},
  flow/.style={->, line width=0.6pt},
  feedback/.style={->, dashed, line width=0.6pt}
]
\node[deterministic] (input)
  {\textbf{Input and control}\\
   $(\Gamma,g)$; atomic claim; pinned versions; budget and log};
\node[deterministic, right=4.5mm of input] (screen)
  {\textbf{Deterministic screening}\\
   source and terminal checks; cheap \textit{autosolve}; pass to gate or enter FAST};
\node[model, right=4.5mm of screen] (fast)
  {\textbf{FAST route}\\
   short hypothesis, witness, binder, theorem, rewrite, or solver plan};
\node[model, right=7.0mm of fast] (deep)
  {\textbf{DEEP route}\\
   targeted retrieval; smallest bridge; decompose and backfill subgaps};

\node[model, below=8.0mm of deep] (loop)
  {\textbf{Model--checker loop}\\
   plan, emit, and revise DSL $\leftrightarrow$ isolated checker with failure evidence};
\node[deterministic, left=4.5mm of loop] (gate)
  {\textbf{Dual acceptance gate}\\
   closed, admit-free root replay; semantic audit: scope, domain, and dependencies};
\node[deterministic, left=4.5mm of gate] (outcome)
  {\textbf{Audited outcome}\\
   accepted certificate or typed diagnosis: invalid, capability, limitation, bug, or blocker};
\node[store, left=4.5mm of outcome] (knowledge)
  {\textbf{Audited reusable knowledge}\\
   proof/reference bank; validated failure experience for later gaps};

\draw[flow] (input) -- (screen);
\draw[flow] (screen) -- (fast);
\draw[flow] (fast) -- node[midway, above=.7mm, fill=white,
  inner xsep=1.0pt, inner ysep=.4pt, font=\small] {stall} (deep);
\draw[flow] (deep) -- (loop);
\draw[flow] (loop) -- (gate);
\draw[flow] (gate) -- (outcome);
\draw[flow] (outcome) -- (knowledge);
\draw[flow] (screen.south) -- ++(0,-4.0mm) -| node[pos=.25, above=.6mm,
  fill=white, inner xsep=1.0pt, inner ysep=.4pt,
  font=\small] {direct pass} (gate.north);
\coordinate (reuseleft) at ($(input.west)+(-6.5mm,0)$);
\draw[feedback] (knowledge.west) -- (reuseleft |- knowledge.west)
  -- (reuseleft) -- (input.west);
\end{tikzpicture}%
}
\caption{Verifier-guided agent baseline.  Darker nodes denote model-directed
work, while dashed arrows denote reuse.  Stalled FAST attempts escalate to
targeted DEEP search; acceptance requires an admit-free root replay and
semantic audit before outcomes can guide later gaps.}
\label{fig:agent-baseline}
\end{figure}

\paragraph{Screening and FAST search.}
Screening performs source and terminal-evidence checks and first tests a cheap
candidate such as \texttt{autosolve all\_hypothesis}.  Unresolved gaps enter
bounded FAST search over short hypothesis, witness, binder, theorem, rewrite,
assertion, and solver routes.  For each attempt, the checker returns the first
failed instruction, compact local evidence, and the residual proof state.  A
retry is justified only when it changes the route or proof state; a repeated
failure signature or exhausted budget triggers escalation rather than another
equivalent solver permutation.

\paragraph{DEEP search and subgaps.}
DEEP search retrieves a small set of relevant theorems, solver cards, recipes,
or verified examples, then identifies the smallest bridge between the current
state and target.  It may organize that bridge as local assertions or recursive
subgaps.  Search-time placeholders can expose those subgaps, but they are never
accepted as evidence: every subgap must be backfilled, and the complete root
program must be replayed with zero admits.

\paragraph{Acceptance and typed diagnoses.}
A machine pass remains provisional.  Acceptance additionally requires a closed,
admit-free root proof and semantic checks of theorem instantiation, quantifier
scope, rewriting, domain conditions, dependencies, and possible inconsistent
assumptions.  Other runs receive evidence-backed
diagnoses such as invalid data, missing theorem or backend capability, known
limitation, bug, or a localized unresolved blocker.  Audited certificates
populate the proof bank, while validated recurring failures populate the
experience bank used for subsequent routing and retrieval.

\paragraph{Audited reuse.}
Atomic claims allow workers to process disjoint gaps concurrently, while
publication into shared state is serialized.  A successful proof enters the
reference bank only after machine and semantic gates pass and duplicate content
is removed.  Repeated failure signatures first become experience candidates;
they influence later routing only after mechanical regression evidence or
review validates the proposed recovery rule.  Retrieval is deliberately
targeted, exposing a small ranked set rather than the complete history.

\FloatBarrier
\subsection{Original-Version Baseline and Reporting Protocol}
\label{app:baseline-reporting}

Treating \texttt{autosolve} as a lightweight automated theorem prover inside
the DSL, we first apply it to all 26,116 instances; it closes 2,838 gaps.  The
remaining 23,278 instances are then processed by our verifier-guided agent
workflow, using Codex as the agent and \texttt{gpt-5.6-sol} as its underlying
model.  Under the same checker and theorem-library snapshot, the workflow
produces accepted DSL proofs for a further 4,889 gaps.  The complete accounting
is shown in Table~\ref{tab:original-baseline-results}.

\begin{table}[!htbp]
\centering
\caption{Checker-accepted certificates for the original ProofGap version.
``Share of all'' is measured over 26,116 gaps; the workflow stage rate is
measured over its 23,278-instance residual input.}
\label{tab:original-baseline-results}
\vspace{4pt}
\PGAppendixTableStyle
\begin{tabularx}{\linewidth}{@{}>{\raggedright\arraybackslash}Xrrrr@{}}
\toprule
\textbf{Stage} & \textbf{Inputs} & \textbf{Newly closed} & \textbf{Share of all} & \textbf{Stage rate} \\
\addlinespace[2pt]
\midrule
Direct \texttt{autosolve} & 26,116 & 2,838 & 10.9\% & 10.9\% \\
\addlinespace[2pt]
Verifier-guided workflow & 23,278 & 4,889 & 18.7\% & 21.0\% \\
\addlinespace[2pt]
\midrule
Combined coverage & 26,116 & 7,727 & 29.6\% & --- \\
\bottomrule
\end{tabularx}
\end{table}

Stage attribution prevents successes found by \texttt{autosolve} from being
credited to the agent workflow.  Operationally, a direct certificate consists
of exactly one nonempty command, after removal of an optional numeric step
prefix, and that command begins with \texttt{autosolve}; every other accepted
program is assigned to the workflow stage.  We report the workflow as a single
baseline rather than attributing completions to its internal FAST or DEEP
routes.

Every accepted DSL program is replayable under the checker and
therefore provides an executable ground-truth solution for the corresponding
gap in the original version of the benchmark.  This is a solution-level notion
of ground truth: multiple distinct DSL programs may validly certify the same
local obligation.  The remaining 18,389 gaps have no packaged certificate and
are not treated as false or unprovable.

\subsection{Certificate Package and Independent Verification}
\label{app:certificate-package}

The final original-version artifact contains one canonical directory for each
of the 26,116 gaps.  Exactly 7,727 directories contain both the gap and an
accepted DSL program, while 18,389 contain the gap only.  Seven secondary
solutions for already represented canonical pairs are excluded, so the
reported count is over unique exercise--gap pairs.  A 26,116-row manifest
records the canonical pair, whether a program is present, provenance, and a
SHA-256 digest for every gap and every available program.  A machine-readable
build summary records the aggregate counts and component digests.  The artifact
also contains the 514-entry theorem snapshot used during replay.

The final artifact includes its packaged verifiers for 64-bit Linux and
Windows together with a Python-standard-library verification driver.  The
driver executes only
cases that contain a DSL program; gap-only cases are counted and skipped.  It
uses isolated temporary settings, result, and log locations for each case and
normalizes UTF-8 byte-order marks, line endings, and blank lines in a temporary
copy, leaving the submitted gap unchanged.  Parallelism is configurable
(default eight workers), as is the per-case timeout (default 120 seconds).  A
full replay is launched with \texttt{python3 check.py}; command-line options can
select the platform, worker count, and timeout.  Running the driver produces
per-case execution and error logs, an aggregate tabular report, and a
machine-readable summary.

A packaged case passes only when all of the following conditions hold:

\begin{enumerate}
  \item the verifier process exits with status zero;
  \item the structured result uses schema
  \texttt{test\_dsl\_structured\_v2};
  \item \texttt{final\_status} is \texttt{finish};
  \item both \texttt{proof\_finished} and \texttt{proof\_verified} are true;
  \item both \texttt{admit\_count} and \texttt{split\_gap\_count} are zero;
  and
  \item standard output contains the terminal marker
  \texttt{dsl proof finished}.
\end{enumerate}

The driver itself exits successfully only when every packaged DSL case passes
all of these checks.  Consequently, a merely parsable program, an unfinished
case branch, an admitted subgap, or a process that exits cleanly without the
structured proof evidence is not classified as passing by the replay driver.

\FloatBarrier
\Needspace{5\baselineskip}
\section{Evaluation Protocol}
\label{app:lean-evaluation}

The Lean evaluation uses two paired target types.  A \emph{proof-gap target}
asks the model to prove one local theorem with the context accumulated from
earlier steps of the reference solution.  Its \emph{parent-exercise target}
asks the model to prove the corresponding complete exercise theorem from the
original premises.  The complete evaluation contains 26,116 proof gaps from
3,015 exercises and their corresponding 3,015 parent-exercise theorems.
The same set of exercises is used for paired gap--exercise and macro-coverage
analyses.  The reference proof is withheld in both settings.

\subsection{Prompt and Verification Interface}
\label{app:lean-prompt-checker}

For a proof-gap target, the prompt contains the original imports, definitions,
helper declarations, statements of preceding gap theorems, and the exact target
statement.  Every gap proof body is replaced by a marker or an omission
comment, and later declarations are excluded.  Earlier theorem statements
remain available because they form part of the local context represented by
the gap.  For a parent-exercise target, the prompt contains the imports,
definitions, and complete exercise statement, but no intermediate gap theorem
or reference proof.  In both cases, the model returns a JSON object whose
\code{proof} field is a complete replacement for the target's \code{by} block.

The returned proof is inserted into the original module, with all other source
text unchanged, and checked in a fresh process.  The pinned environment uses
Lean \code{v4.29.0-rc6} and Mathlib revision
\code{5c8398df\allowbreak{}528176d9\allowbreak{}c87ccd92\allowbreak{}26ba8f7c\allowbreak{}8852d59c}.  A candidate is accepted only
if the reconstructed module compiles and a recursive \code{assert\_no\_sorry}
check confirms that the target has no dependency on \code{sorryAx}.  The
checker also rejects proof escapes and declarations that could change the
trusted environment, including \code{sorry}, \code{admit}, \code{axiom},
\code{constant}, and \code{run\_tac}.  Candidate checks are isolated so that
generated files and process state cannot be reused across targets.

\subsection{Models and Candidate Budget}
\label{app:model-budget}

The full paired evaluation uses Goedel-Prover-V2-8B,
DeepSeek-Prover-V2-7B, and Kimina-Prover-Distill-8B.  ProofGap-280 additionally
includes Kimi-K3, DeepSeek-V4-Pro, GLM-5.2, Claude-Opus-4.8, and GPT-5.6-sol.
Each model receives the same target-specific prompt interface and generates
eight candidate proofs per target.  Each generation is capped at 8,192 output
tokens.  We use temperature \(0.6\) and nucleus-sampling parameter
\(\mathrm{top}\_p=0.95\).  The comparison is therefore controlled at the
\emph{target} level: one proof gap and one complete exercise theorem each
receive eight candidates.  It is not an exercise-level compute comparison,
because an exercise containing multiple gaps induces multiple independently
evaluated local targets.

\subsection{Pass@\(k\) and Exercise-Level Coverage}
\label{app:evaluation-metrics}

For fixed-\(n\) runs, let \(T\) be the number of targets, \(n=8\) the
number of checked candidates per target, and \(c_i\) the number of accepted
candidates for target \(i\). For \(k\leq n\), we use the standard unbiased
estimator of \citet{chen2021evaluating}
\[
  \widehat{\mathrm{Pass@}k}
  =
  \frac{1}{T}\sum_{i=1}^{T}
  \left(
    1-\frac{\binom{n-c_i}{k}}{\binom{n}{k}}
  \right),
\]
with the ratio interpreted as zero when \(n-c_i<k\).  At \(k=8\), this
reduces exactly to the fraction of targets with at least one accepted
candidate.  All reported Lean evaluation runs have eight checked candidates
per target and use this estimator.  Malformed, timed-out, prohibited, or
noncompiling returned candidates count as failures.

For the full proof-gap set, micro averaging assigns equal weight to each gap.
The exercise-level diagnostic analyses use Pass@8. Let \(N\) be the number of
paired exercises, \(n_i\) the number of gaps in exercise \(i\), and
\(y_{ij}^{(8)}\in\{0,1\}\) indicate whether at least one checked candidate
solves its \(j\)-th gap under the eight-candidate budget. At this budget, micro-averaged gap
performance and macro-averaged exercise coverage are
\[
  \mathrm{GapPass}_{\mathrm{micro}}@8
  =
  \frac{\sum_{i=1}^{N}\sum_{j=1}^{n_i}y_{ij}^{(8)}}
       {\sum_{i=1}^{N}n_i},
  \qquad
  \mathrm{GapCov}_{\mathrm{macro}}@8
  =
  \frac{1}{N}\sum_{i=1}^{N}
  \frac{1}{n_i}\sum_{j=1}^{n_i}y_{ij}^{(8)}.
\]
Micro averaging weights all gaps equally; macro coverage first computes the
fraction of solved gaps within each exercise and then weights all exercises
equally.
The reported exercise-level means and conditional analyses use all 3,015
paired exercises.

\FloatBarrier
\Needspace{5\baselineskip}
\section{ProofGap-280 Construction}
\label{app:proofgap-280}

ProofGap-280 is a reduced-cost, balanced subset for cross-model comparison.
Its construction jointly controls mathematical topic and the length of the
Lean reference proof.  Reference-proof length is used only as reproducible
sampling metadata: the reference proof itself, its length, and its hash are
not included in the model prompt, and line count is not assumed to be a
complete measure of mathematical difficulty.

\subsection{Length Strata and Topic Mapping}
\label{app:proofgap-280-strata}

The benchmark catalog parses the exact \code{by} proof body of each gap and
counts its physical lines after removing trailing whitespace.  The four
strata are
\[
\begin{array}{lll}
  \textsc{Tiny}:   & 1\text{--}2   & \text{gold-proof lines},\\
  \textsc{Short}:  & 3\text{--}5   & \text{gold-proof lines},\\
  \textsc{Medium}: & 6\text{--}15  & \text{gold-proof lines},\\
  \textsc{Long}:   & 16\text{ or more} & \text{gold-proof lines}.
\end{array}
\]
The seven topics follow the source-area mapping used in the main paper:
functions, limits, and continuity; single-variable differentiation;
indefinite and definite integrals; series; multivariable differentiation;
parameter integrals; and multiple, line, and surface integrals.

\subsection{Balanced Sampling Procedure}
\label{app:proofgap-280-sampling}

We form the Cartesian product of seven topics and four length strata, producing
28 sampling cells.  Each cell contributes exactly ten gaps, giving 40 gaps per
topic, 70 per length stratum, and 280 in total.  Within each cell, eligible
catalog entries are shuffled by Python's deterministic
\code{random.Random(20260727)} generator and selected subject to at most one
gap from each base exercise number.  The latter rule also prevents variants
such as two separately indexed subquestions of the same numbered exercise from
both entering the subset.  Consequently, the 280 gaps have 280 distinct
parent-exercise targets.  After selection, rows are sorted by exercise number,
exercise variant, and gap number to give a stable evaluation order.

\Needspace{22\baselineskip}
\begin{table}[!htbp]
\centering
\caption{ProofGap-280 allocation across the 28 topic-by-length cells.}
\label{tab:proofgap-280-allocation}
\vspace{4pt}
\PGAppendixTableStyle
\begin{tabularx}{\linewidth}{@{}>{\raggedright\arraybackslash}Xrrrrr@{}}
\toprule
\textbf{Topic} & \textbf{Tiny} & \textbf{Short} & \textbf{Medium} & \textbf{Long} & \textbf{Total} \\
\addlinespace[2pt]
\midrule
Functions, limits, and continuity & 10 & 10 & 10 & 10 & 40 \\
\addlinespace[2pt]
Single-variable differentiation & 10 & 10 & 10 & 10 & 40 \\
\addlinespace[2pt]
Indefinite and definite integrals & 10 & 10 & 10 & 10 & 40 \\
\addlinespace[2pt]
Series & 10 & 10 & 10 & 10 & 40 \\
\addlinespace[2pt]
Multivariable differentiation & 10 & 10 & 10 & 10 & 40 \\
\addlinespace[2pt]
Parameter integrals & 10 & 10 & 10 & 10 & 40 \\
\addlinespace[2pt]
Multiple, line, and surface integrals & 10 & 10 & 10 & 10 & 40 \\
\addlinespace[2pt]
\midrule
Total & 70 & 70 & 70 & 70 & 280 \\
\bottomrule
\end{tabularx}
\end{table}

The released subset manifest records the seed, strategy, cell counts, stable
subset position, exercise identifier, source topic, length stratum, reference
proof line count, gap identifier, and paired original-exercise identifier for
every selected row.  It therefore permits exact regeneration of the subset
and direct alignment of every gap result with its parent-exercise result.

\clearpage
\FloatBarrier
\Needspace{5\baselineskip}
\section{Additional Results and Analysis}
\label{app:proofgap-280-results}

\subsection{Full-Set Pass@\(k\)}
\label{app:full-passk}

Table~\ref{tab:full-passk-results} gives the complete Pass@1, Pass@3, and
Pass@8 results for the three specialized provers.  The proof-gap columns use
all 26,116 local targets, and the parent-exercise columns use their 3,015
corresponding complete theorems.

\begin{table}[!htbp]
\centering
\caption{Full-set Pass@\(k\) (\%) for proof gaps and complete parent-exercise
theorems.}
\label{tab:full-passk-results}
\vspace{4pt}
\PGAppendixTableStyle
\begin{tabular*}{\linewidth}{@{\extracolsep{\fill}}lrrrrrr@{}}
\toprule
& \multicolumn{3}{c}{All proof gaps} &
\multicolumn{3}{c}{Parent exercises} \\
\cmidrule(lr){2-4}\cmidrule(l){5-7}
Model & Pass@1 & Pass@3 & Pass@8 & Pass@1 & Pass@3 & Pass@8 \\
\midrule
Goedel-Prover-V2-8B      & 26.60 & 32.85 & 35.39 & 2.17 & 3.36 & 4.28 \\
DeepSeek-Prover-V2-7B    & 26.12 & 31.84 & 33.27
                         & 1.92 & 2.67 & 3.32 \\
Kimina-Prover-Distill-8B & 25.56 & 30.77 & 31.99 & 1.74 & 2.42 & 2.92 \\
\bottomrule
\end{tabular*}
\end{table}

The standard-tactic union solves 2,925 gaps, or 11.20\% of the full set.
Removing these leaves 23,191 tactic-unsolved gaps (88.80\%).  As shown in
Table~\ref{tab:hard-gap-results}, the learned provers retain substantial
performance on this residual set; their results are not explained only by
obligations already solved by the evaluated tactic portfolio.

\begin{table}[!htbp]
\centering
\caption{
Solve rates of standard Lean tactics on 26,116 proof gaps.
The union succeeds if any evaluated tactic succeeds.
}
\label{tab:tactic-baselines}
\vspace{4pt}
\PGAppendixTableStyle
\begin{tabular*}{.66\linewidth}{@{\extracolsep{\fill}}lr@{}}
\toprule
\textbf{Tactic} & \textbf{Solve rate (\%)} \\
\midrule
\texttt{assumption} & 1.78 \\
\texttt{simp}       & 1.88 \\
\texttt{aesop}      & 9.67 \\
\texttt{norm\_num}  & 2.45 \\
\texttt{linarith}   & 2.75 \\
\texttt{nlinarith}  & 2.85 \\
\texttt{ring}       & 1.08 \\
\texttt{omega}      & 0.17 \\
\midrule
\textbf{Union}      & \textbf{11.20} \\
\bottomrule
\end{tabular*}
\end{table}

\begin{table}[!htbp]
\centering
\caption{Performance (\%) on the 23,191 gaps not solved by the standard-tactic
union.}
\label{tab:hard-gap-results}
\vspace{4pt}
\PGAppendixTableStyle
\begin{tabular*}{\linewidth}{@{\extracolsep{\fill}}lrrr@{}}
\toprule
\textbf{Model} & \textbf{Pass@1} & \textbf{Pass@8} & \textbf{Solved at Pass@8} \\
\midrule
Goedel-Prover-V2-8B      & 18.61 & 27.48 & 6,374 \\
DeepSeek-Prover-V2-7B    & 17.81 & 25.16 & 5,834 \\
Kimina-Prover-Distill-8B & 17.07 & 23.57 & 5,467 \\
\bottomrule
\end{tabular*}
\end{table}

\clearpage
\subsection{ProofGap-280 Breakdown}
\label{app:proofgap-280-breakdown}

The main paper reports overall Pass@1, Pass@3, and Pass@8 for all models on
ProofGap-280.  The following tables give Pass@8 breakdowns for the five
general-purpose models.  Each entry in
Tables~\ref{tab:proofgap-280-topic-results} and
\ref{tab:proofgap-280-length-results} reports
\emph{proof gap / parent exercise}.  Topic cells have denominator 40 and
length cells have denominator 70.

\Needspace{42\baselineskip}
\begin{table}[!htbp]
\centering
\caption{ProofGap-280 Pass@8 (\%) by mathematical topic.  Each entry is
proof-gap / parent-exercise performance.}
\label{tab:proofgap-280-topic-results}
\vspace{4pt}
\PGAppendixTableStyle
\begin{tabular*}{\linewidth}{@{\extracolsep{\fill}}lcccc@{}}
\toprule
\textbf{Model} & \textbf{\begin{tabular}[c]{@{}c@{}}Functions\\\& limits\end{tabular}} & \textbf{\begin{tabular}[c]{@{}c@{}}Single-var.\\diff.\end{tabular}} & \textbf{\begin{tabular}[c]{@{}c@{}}Indef./def.\\integrals\end{tabular}} & \textbf{Series} \\
\midrule
GPT-5.6-sol & \PGAppendixPair{95.0}{55.0} & \PGAppendixPair{90.0}{20.0} & \PGAppendixPair{90.0}{2.5} & \PGAppendixPair{90.0}{17.5} \\
Claude-Opus-4.8 & \PGAppendixPair{87.5}{37.5} & \PGAppendixPair{87.5}{17.5} & \PGAppendixPair{92.5}{2.5} & \PGAppendixPair{92.5}{10.0} \\
Kimi-K3 & \PGAppendixPair{85.0}{27.5} & \PGAppendixPair{80.0}{12.5} & \PGAppendixPair{90.0}{0.0} & \PGAppendixPair{80.0}{5.0} \\
DeepSeek-V4-Pro & \PGAppendixPair{65.0}{10.0} & \PGAppendixPair{57.5}{5.0} & \PGAppendixPair{77.5}{0.0} & \PGAppendixPair{65.0}{5.0} \\
GLM-5.2 & \PGAppendixPair{65.0}{5.0} & \PGAppendixPair{57.5}{2.5} & \PGAppendixPair{72.5}{0.0} & \PGAppendixPair{67.5}{0.0} \\
\bottomrule
\end{tabular*}
\par\vspace{10pt}
\begin{tabular*}{\linewidth}{@{\extracolsep{\fill}}lcccc@{}}
\toprule
\textbf{Model} & \textbf{\begin{tabular}[c]{@{}c@{}}Multivar.\\diff.\end{tabular}} & \textbf{\begin{tabular}[c]{@{}c@{}}Parameter\\integrals\end{tabular}} & \textbf{\begin{tabular}[c]{@{}c@{}}Multiple/line/\\surface integrals\end{tabular}} & \textbf{Overall} \\
\midrule
GPT-5.6-sol & \PGAppendixPair{90.0}{25.0} & \PGAppendixPair{77.5}{7.5} & \PGAppendixPair{82.5}{22.5} & \PGAppendixPair{87.9}{21.4} \\
Claude-Opus-4.8 & \PGAppendixPair{90.0}{10.0} & \PGAppendixPair{82.5}{2.5} & \PGAppendixPair{82.5}{17.5} & \PGAppendixPair{87.9}{13.9} \\
Kimi-K3 & \PGAppendixPair{82.5}{5.0} & \PGAppendixPair{75.0}{2.5} & \PGAppendixPair{75.0}{10.0} & \PGAppendixPair{81.1}{8.9} \\
DeepSeek-V4-Pro & \PGAppendixPair{65.0}{5.0} & \PGAppendixPair{67.5}{2.5} & \PGAppendixPair{75.0}{7.5} & \PGAppendixPair{67.5}{5.0} \\
GLM-5.2 & \PGAppendixPair{60.0}{5.0} & \PGAppendixPair{65.0}{2.5} & \PGAppendixPair{67.5}{7.5} & \PGAppendixPair{65.0}{3.2} \\
\bottomrule
\end{tabular*}
\end{table}

\begin{table}[!htbp]
\centering
\caption{ProofGap-280 Pass@8 (\%) by the reference proof length of the
selected gap.  Each entry is proof-gap / parent-exercise performance.  The
parent-exercise column is grouped by its paired gap's stratum, not by the
length of a parent-exercise reference proof.}
\label{tab:proofgap-280-length-results}
\vspace{4pt}
\PGAppendixTableStyle
\begin{tabular*}{\linewidth}{@{\extracolsep{\fill}}lccccc@{}}
\toprule
\textbf{Model} & \textbf{Tiny} & \textbf{Short} & \textbf{Medium} & \textbf{Long} & \textbf{Overall} \\
\midrule
GPT-5.6-sol
& \PGAppendixPair{100.0}{22.9} & \PGAppendixPair{98.6}{24.3} & \PGAppendixPair{92.9}{22.9} & \PGAppendixPair{60.0}{15.7} & \PGAppendixPair{87.9}{21.4} \\
Claude-Opus-4.8
& \PGAppendixPair{98.6}{12.9} & \PGAppendixPair{98.6}{18.6} & \PGAppendixPair{88.6}{14.3} & \PGAppendixPair{65.7}{10.0} & \PGAppendixPair{87.9}{13.9} \\
Kimi-K3
& \PGAppendixPair{98.6}{8.6} & \PGAppendixPair{98.6}{11.4} & \PGAppendixPair{92.9}{10.0} & \PGAppendixPair{34.3}{5.7} & \PGAppendixPair{81.1}{8.9} \\
DeepSeek-V4-Pro
& \PGAppendixPair{100.0}{7.1} & \PGAppendixPair{91.4}{8.6} & \PGAppendixPair{65.7}{2.9} & \PGAppendixPair{12.9}{1.4} & \PGAppendixPair{67.5}{5.0} \\
GLM-5.2
& \PGAppendixPair{97.1}{4.3} & \PGAppendixPair{87.1}{5.7} & \PGAppendixPair{64.3}{1.4} & \PGAppendixPair{11.4}{1.4} & \PGAppendixPair{65.0}{3.2} \\
\bottomrule
\end{tabular*}
\end{table}
\FloatBarrier

\paragraph{Length sensitivity.}
All five models solve at least 97.1\% of Tiny gaps, whereas Long-gap Pass@8
ranges from 65.7\% for Claude-Opus-4.8 to 11.4\% for GLM-5.2.  Much of the
cross-model separation is therefore concentrated in longer local derivations.
Reference-proof line count remains a coverage proxy rather than a complete
measure of intrinsic difficulty.

\paragraph{Local versus end-to-end reasoning.}
GPT-5.6-sol and Claude-Opus-4.8 tie at 87.9\% on gaps but reach 21.4\% and
13.9\%, respectively, on the paired exercises.  Even in the Tiny stratum,
parent-exercise Pass@8 is only 4.3--22.9\%.  The parent rows are grouped by the
paired gap's length, not by parent-theorem proof length.

\paragraph{Topic sensitivity.}
Parent exercises are easiest in functions and limits for every model.
Indefinite and definite integral gaps reach 72.5--92.5\%, but the paired
exercise score is 2.5\% for GPT-5.6-sol and Claude-Opus-4.8 and zero for the
other models; parameter integrals are also difficult end to end.  Local
success therefore leaves substantial work in lemma selection, analytic side
conditions, and proof composition.
\FloatBarrier
\endgroup

\end{document}